\documentclass[letterpaper,twocolumn,10pt]{article}
\usepackage{usenix}

\usepackage{tikz}
\usepackage{amsmath}

\usepackage{adjustbox}
\usepackage{amsmath} 
\usepackage{amssymb}
\usepackage{amsthm}
\usepackage{appendix}

\usepackage{caption}
\usepackage{enumitem}
\usepackage{multirow}
\usepackage{subcaption}
\usepackage{xspace}

\newcommand{\diff}[1]{\textcolor{blue}{#1}}

\newcommand*\circled[1]{\tikz[baseline=(char.base)]{
            \node[shape=circle,draw,fill=black,text=white,inner sep=0.5pt] (char) {\textbf{#1}};}}

\newtheorem{theorem}{Theorem}
\newtheorem{lemma}{Lemma}

\newtheorem{assumption}{Assumption}

\newcommand{\system}{\textsc{GeoShield}\xspace}
\newcommand{\rebound}{\textsc{Rebound}\xspace}
\newcommand{\roborebound}{\textsc{RoboRebound}\xspace}

\newcommand{\pbft}{\texttt{PBFT}\xspace}
\newcommand{\Zyzzyva}{\texttt{Zyzzyva}\xspace}
\newcommand{\ZyzzyvaIdeal}{\texttt{Zyzzyva-ideal}\xspace}
\newcommand{\ZyzzyvaOmission}{\texttt{Zyzzyva-omission}\xspace}
\newcommand{\msptp}{\texttt{MS-PTP}\xspace}
\newcommand{\pistis}{\texttt{PISTIS}\xspace}

\newcommand{\bsection}[1]{\noindent\textbf{#1.}}

\newcommand{\Dinter}{\ensuremath{\Delta^{\mathsf{d}}}\xspace} 
\newcommand{\pnorm}{\ensuremath{P_{\mathsf{norm}}}\xspace} 
\newcommand{\Dintra}{\ensuremath{\Delta_{\mathsf{intra}}}\xspace} 
\newcommand{\tintra}{\ensuremath{D_{\mathsf{intra}}}\xspace} 
\newcommand{\Dsyn}{\ensuremath{\Delta_{\mathsf{syn}}}\xspace} 
\newcommand{\thb}{\ensuremath{E_{\mathsf{hb}}}\xspace} 
\newcommand{\Dhb}{\ensuremath{\Delta_{\mathsf{hb}}}\xspace} 
\newcommand{\tprop}{\ensuremath{E_{\mathsf{prop}}}\xspace} 
\newcommand{\Dprop}{\ensuremath{\Delta_{\mathsf{prop}}}\xspace} 

\newcommand{\tinterval}{\ensuremath{R_{\mathsf{hb}}}\xspace} 
\newcommand{\tsendn}{\ensuremath{t_{n}}\xspace} 
\newcommand{\tdetect}{\ensuremath{t_{\mathsf{det}}}\xspace} 
\newcommand{\tearly}{\ensuremath{\Delta_{\mathsf{early}}}\xspace} 
\newcommand{\thbtimeout}{\ensuremath{D_{\mathsf{t/o}}}\xspace} 
\newcommand{\taccept}{\ensuremath{D_{\mathsf{acc,n}}}\xspace}
\newcommand{\tdecide}{\ensuremath{D_n}\xspace}

\newcommand{\tpocgap}{\ensuremath{D_{\mathsf{gap}}^{\mathsf{poc}}}}

\newcommand{\spenaty}{\ensuremath{s_{\mathsf{pen}}}\xspace} 
\newcommand{\saward}{\ensuremath{s_{\mathsf{awd}}}\xspace} 

\newcommand{\recoverybound}{\ensuremath{D_{\max}^{\mathsf{rec}}}}
\newcommand{\recoveryboundintra}{\ensuremath{D_{\mathsf{intra}}^{\mathsf{rec}}}}

\begin{document}

\date{}

\title{\Large \bf {\system}: Byzantine Fault Detection and Recovery for Geo-Distributed \\ Real-Time Cyber-Physical Systems
}
\author{{\rm Yifan Cai} \\
University of Pennsylvania
\and
{\rm Linh Thi Xuan Phan} \\
University of Pennsylvania
}

\maketitle

\begin{abstract}
Large-scale cyber-physical systems (CPS), such as railway control systems and smart grids, consist of geographically distributed subsystems that are connected via unreliable, asynchronous inter-region networks.
Their scale and distribution make them especially vulnerable to faults and attacks. Unfortunately, existing fault-tolerant methods either consume excessive resources or provide only eventual guarantees, making them unsuitable for real-time resource-constrained CPS. 

We present \system,  a resource-efficient solution for defending geo-distributed CPS against Byzantine faults. \system leverages the property that CPS are designed to tolerate brief disruptions and maintain safety, as long as they recover (i.e., resume normal operations or transition to a safe mode) within a bounded amount of time following a fault. Instead of masking faults, it \emph{detects} them and \emph{recovers} the system within {\em bounded time}, thus guaranteeing safety with much fewer resources. \system introduces protocols for Byzantine fault-resilient network measurement and inter-region omission fault detection that proactively detect malicious message delays, along with recovery mechanisms that guarantee timely recovery while maximizing operational robustness. It is the first bounded-time recovery solution that operates effectively under unreliable networks without relying on trusted hardware.
Evaluations using real-world case studies show that it significantly outperforms existing methods in both effectiveness and resource efficiency.

\end{abstract}

\section{Introduction}\label{sec:intro}
\begin{figure*}[htbp]
    \centering
    \begin{minipage}{0.35\textwidth} 
        \centering
        \includegraphics[width=.9\linewidth]{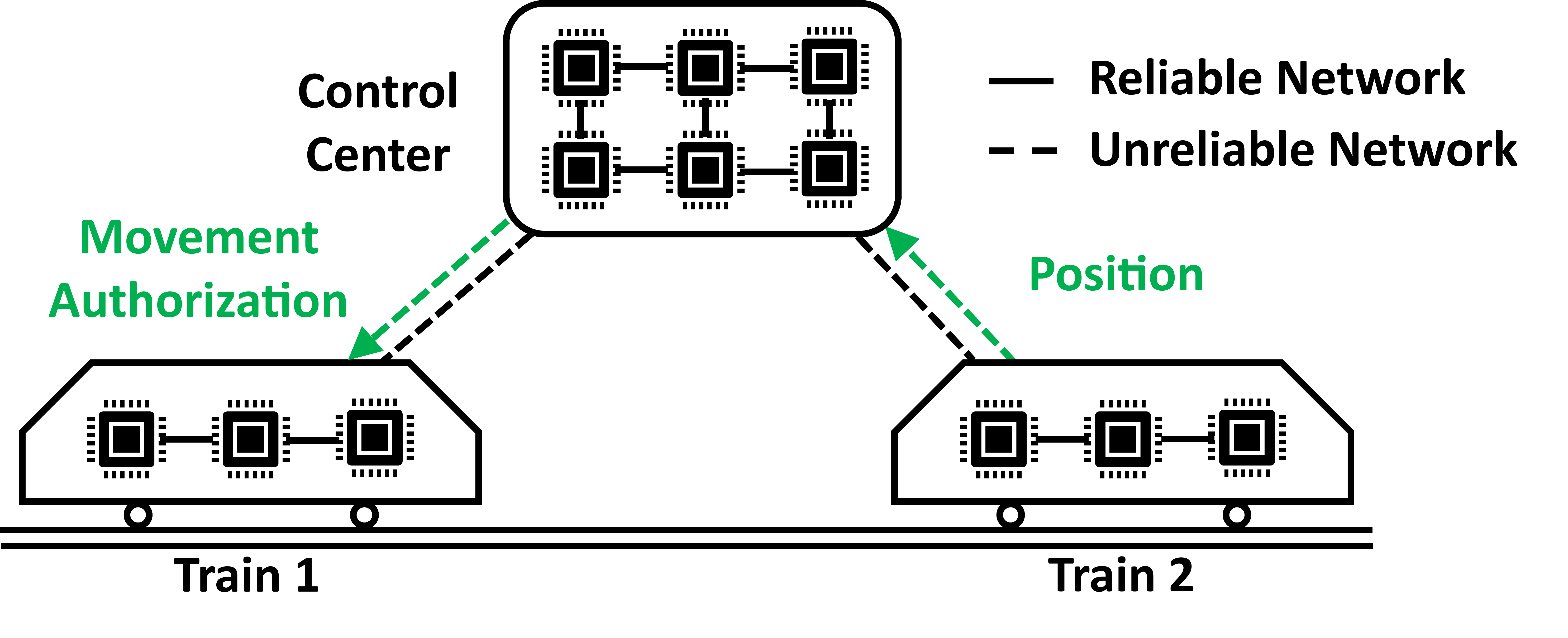} 
        \caption{A railway control system~\cite{hollysys-brochure}.}
        \label{fig:train-example}
    \end{minipage} 
    \hspace{1ex}
    \begin{minipage}{0.63\textwidth}
        \centering
        \small
        \scalebox{0.8}{
        \begin{tabular}{|c|c|c|c|c|c|}
        \hline
            \textbf{Solutions} &
              \textbf{\#Replicas} &
              \textbf{\begin{tabular}[c]{@{}c@{}}Trusted\\ HW\end{tabular}} &
              \textbf{\begin{tabular}[c]{@{}c@{}}Assume\\ Synchrony\end{tabular}} &
              \textbf{\begin{tabular}[c]{@{}c@{}}Timing\\ Guarantees\end{tabular}} &
              \textbf{\begin{tabular}[c]{@{}c@{}}Operational\\ Robustness\end{tabular}} \\ \hline
            \rebound \cite{gandhi2021rebound}                                 & $\geq$ {\bf f + 1} & {\bf No} & Yes         & \textbf{Yes} & \textbf{Proactive / Effective} \\ \hline
            PBFT~\cite{castro1999practical} / Zyzzyva~\cite{kotla2007zyzzyva} & $\geq$ 3f + 1      & {\bf No} & \textbf{No} & No           & Reactive / Limited             \\ \hline
            PISTIS~\cite{DBLP:journals/tpds/KozhayaDRV21}                     & $\geq$ 3f + 1      & {\bf No} & \textbf{No} & \textbf{Yes} & Reactive / Limited             \\ \hline
            \roborebound~\cite{gandhi2025roborebound}                       & $\geq$ {\bf f + 1}  & Yes &   \textbf{No} & \textbf{Yes} & Reactive / Limited             \\ \hline
            \system                                                           & $\geq$ {\bf f + 1} & {\bf No} &  \textbf{No} & \textbf{Yes} & \textbf{Proactive / Effective} \\ \hline
        \end{tabular}
        }
        \captionof{table}{\system vs. existing solutions.}
        \label{tbl:comparison}
    \end{minipage}
\end{figure*}

\noindent Large-scale cyber-physical systems (CPS)---such as traffic control systems~\cite{hollysys-brochure,etcs3-spec}, power grids~\cite{bommareddy2022real,DBLP:journals/pieee/HuangPDOLGZ24,kim2025revisiting}, and industrial control systems~\cite{DBLP:conf/raid/Zhang00M22}---are commonly geo-distributed. A railway control system~\cite{hollysys-brochure}, for example, consists of subsystems in different regions, such as onboard a train or in a control center (see Fig.~\ref{fig:train-example}). These subsystems must communicate to ensure safe coordination and the shared use of tracks across large distances. 
Within a region, nodes are connected via a synchronous and reliable network, using technologies such as Time-Triggered Ethernet~\cite{tte-tttech} or FlexRay~\cite{flexray} to achieve bounded network latency and zero packet loss.
In contrast, inter-region networks, such as GSM-R~\cite{etcs3-spec} or LTE-R~\cite{lte-cloud-huawei,he2016high} used in railway systems, are asynchronous and unreliable.

As CPS interact with the real world, {\em ensuring correctness and timing guarantees is critical}. For instance, control messages sent from a control center to a train must be both {\em accurate} and {\em timely} to prevent accidents. However, achieving these guarantees in geo-distributed CPS is highly challenging due to their vulnerability to faults that arise from severe weather, hardware failure, and malicious attacks. 

\bsection{Motivating case study: The incident in Wenzhou, China}
The incident in Wenzhou, China, in 2011~\cite{bbc-wenzhou}, illustrates the prevalence and severity of Byzantine faults in geo-distributed CPS. In this incident, two trains were moving on the same track when lightning broke the communication nodes of the first (front) train. This train's driver managed to take over and moved the train at a limited speed of 20 km/h. Due to unreliable inter-region network communication, the control center did not get the first train's updated status.  
Further, because of a Byzantine fault in the control unit, the second train was authorized to move onto the occupied track. The two trains crashed into one another, causing 40 deaths and over 192 injuries. Similar faults or cyber attacks on large-scale CPS have also led to pollution~\cite{nyt-florida}, power outages~\cite{cnn-ukraine}, and fatalities~\cite{ap-india}.

\noindent{\bf Approach.} A standard way to handle Byzantine faults is Byzantine Fault Tolerance (BFT). However, BFT  suffers from high resource consumption: in asynchronous environments (e.g., inter-region networks), it requires $3f+1$ replica nodes to tolerate $f$ faulty nodes, which is often too costly for CPS due to their stringent resource constraints. 
Unlike BFT, which masks faults, we {\em detect} faults and {\em recover} from them {\em within bounded time}. This approach, called {\em bounded-time recovery} (BTR)~\cite{Gandhi2020BoundedtimeRF}, exploits two properties of CPS: 1) they have physical properties (e.g., inertia or thermal capacity) that limit the rate at which state changes can occur, and 2) their control tasks are designed to tolerate some amount of noise or missing inputs without affecting safety~\cite{soudbakhsh2013co}. Thus, CPS can tolerate short periods of arbitrary behavior---the system remains safe as long as it recovers within a bounded amount of time (e.g., 500~ms for building control~\cite{morari-cdc2010} and 2~s for underfrequency load shedding in power grids~\cite{nerc2011reliability}). BTR substantially reduces resource requirements (e.g., only $f+1$ replicas are needed to detect $f$ faults) while preserving 
safety guarantees.

\noindent{\bf Existing BTR methods.} There exist two BTR solutions for Byzantine faults. The first, \rebound~\cite{gandhi2021rebound}, assumes a {\em fully} synchronous and reliable network and thus cannot be directly applied to our setting. When a fault is detected, correct nodes must disseminate evidence to all nodes to enable recovery. If these messages are delayed or dropped by inter-region networks, the system may fail to recover in time, leading to \emph{safety} violations. Recently, \roborebound~\cite{gandhi2025roborebound} extended \rebound to multi-robot systems with asynchronous, unreliable networks. However, \roborebound requires trusted hardware, which is difficult to deploy in large-scale CPS environments. Achieving bounded-time recovery without trusted hardware is challenging in unreliable networks as it is difficult to distinguish between messages delayed or dropped by malicious nodes and those affected by poor network conditions. Malicious nodes can exploit this ambiguity to delay recovery or force correct nodes into safe mode by repeatedly dropping or delaying messages. Since the system operates at reduced functionality in safe mode, we must ensure timely recovery while minimizing unnecessary safe-mode activations to maintain \emph{operational robustness}.

\noindent{\bf Contributions.} We present \system, a novel BTR solution for handling Byzantine faults in geo-distributed CPS. Unlike other distributed systems, CPS often use dedicated networks or resource reservations~\cite{nokia-backbone,hollysys-brochure} with careful scheduling to ensure sufficient network bandwidth and avoid congestion. Thus, although inter-region networks are unreliable and latency varies over time, there exists a {\em probabilistic bound on jitter} within  short intervals. Specifically, if two messages are sent from the same region to another region within a small time window $\epsilon_t$, their arrival times will differ by at most a small $\Delta_t$ with high probability. This enables us to distinguish intentional delays from normal ones: if messages $m$ and $m'$ are sent nearly simultaneously but $m$ arrives significantly later than $m'$, it is highly likely that $m$ was intentionally delayed.  

Based on this insight, we develop a \emph{Byzantine-resilient network measurement protocol} for estimating inter-region network latency in the presence of Byzantine nodes. Our protocol ensures that correct nodes will {\em agree} on a probabilistic maximum value of current network latency, and that the agreed value is within a small difference of the actual probabilistic maximum, even if Byzantine nodes behave arbitrarily. This agreement---which has not been achieved in prior work---is essential for consistent omission fault detection in \system and for normal operations in CPS applications that rely on current network latency as input~\cite{hu2021lars,ateya2019energy}. 

Leveraging this estimation, we design a \emph{distributed timeliness governing system (TGS)} that \emph{proactively} mitigates the impact of delayed or dropped messages. 
TGS maintains operational robustness by {\em bounding} (intentional) delay or loss caused by a faulty node, in both the long term and the short term. 
Once a node reaches its delay or drop limit, it is temporarily excluded from inter-region communication, thus reducing its chance of causing the system to switch to safe mode unnecessarily. Our evaluation shows that this proactive approach outperforms reactive mechanisms in prior work~\cite{DBLP:journals/tpds/KozhayaDRV21}. 

Finally, we introduce fault detection and recovery mechanisms that guarantee the system will recover---i.e., resume normal operation or switch to safe mode---within a bounded time after a fault occurs, even in the presence of inter-region benign message drops/delays and Byzantine behavior by faulty nodes. Importantly, our {\em safety guarantee holds even if the probabilistic bounded jitter is violated at runtime.} 


In summary, our contributions are: \circled{1} a measurement study showing the probabilistic bounded jitter property of inter-region latency; \circled{2}  a Byzantine fault-resilient network measurement protocol ensuring agreement among correct nodes on a probabilistic maximum latency close to the actual probabilistic maximum; \circled{3} mechanisms to detect and recover within bounded time from inter-region Byzantine faults; 
and \circled{4} case studies of railway control and smart grid systems demonstrating our technique's utility.  

To the best of our knowledge, \system is the first BTR solution for geo-distributed CPS that guarantees safety and optimizes operational robustness under Byzantine faults and unreliable networks, {\em without trusted hardware} (see Table~\ref{tbl:comparison}).


\bsection{Limitations}
\system primarily addresses Byzantine attacks on compute nodes. While it can be integrated with existing methods for mitigating attacks against sensors and actuators~\cite{zhou2022security,DBLP:conf/rtas/ZhangSLLCKSL23}, it does not specifically handle DoS or other network attacks. However, we discuss the implications of such attacks on our assumption and security properties in Sec.~\ref{sec:discussion}.
\section{Models and Approach Overview}

\subsection{Standard properties of CPS}\label{subsec:sys-model}
\noindent As illustrated in Fig.~\ref{fig:train-example}, a geo-distributed CPS is composed of multiple regions, $\{R_1, R_2, \dots, R_n\}$, with each $R_i$ consisting of $n_i$ nodes that execute a set of control tasks. 
Since the CPS interacts with the physical world through sensors and actuators, all tasks and software are designed to have {\em bounded execution time}; their worst-case execution times (WCETs) can be obtained using existing WCET analysis tools~\cite{dardaillon2019reconciling,falk2011wcet}. 
Additionally, since sensing and actuation are performed at fixed frequencies, tasks are typically invoked periodically: at each invocation, they take either sensor data or outputs from upstream tasks as inputs, perform computation, and produce results used by actuators or downstream tasks. We call each instance (invocation) of a task a {\em job}. 
For timing guarantees, nodes' clocks are synchronized within a small $\Dsyn$~\cite{d2017time}.  
As is standard in prior work on Byzantine faults and on CPS, we assume all tasks and software are {\em deterministic}~\cite{castro1999practical,gandhi2025roborebound,DBLP:conf/podc/YinMRGA19}. 

The intra-region networks are synchronous and reliable~\cite{hollysys-brochure,tte-tttech,flexray}, i.e., there exists {\em a known maximum intra-region network latency} $\tintra$ between any two nodes in the same region. However, the inter-region networks are asynchronous and unreliable. 
To maintain safety in cases of poor network conditions, each region is typically designed to have a special {\em safe mode}  that it should switch to if any of its tasks has received no input from an upstream region for a certain timeout~\cite{etcs3-spec,finnish-etcs-values,itea-etcs-values}. 
 As we focus on detection and recovery, we assume an existing method (e.g.,~\cite{Gandhi2020BoundedtimeRF,gandhi2021rebound}) is used to compute a schedule for the system that ensures timing guarantees.

\subsection{Threat model}
\noindent We consider the Byzantine fault model, in which faulty nodes may behave arbitrarily---including dropping or delaying messages, sending incorrect outputs, or colluding with other Byzantine nodes both within and across regions. 
A node may commit one or both types of faults: an \emph{omission fault} occurs when a node fails to deliver a message by its expected deadline, and a \emph{commission fault} occurs when a node sends an incorrect message or one that should not have been sent. (Each node is first checked for omission faults---if a message arrives too late, it is discarded; otherwise, it is evaluated for commission faults.) A fault is an {\em intra-region fault} if both the sender and receiver are in the same region, and an {\em inter-region fault} otherwise. 
As in prior work~\cite{gandhi2021rebound}, we define a node to be \emph{correct} if its external \emph{observable} behaviors---those visible to other nodes---are consistent with its expected functionality. A compromised node may pretend to be correct; however, without a trusted mechanism to inspect its internal state, such a fault is impossible to detect~\cite{haeberlen-osdi09}. Since the faulty node does what correct nodes would do, it introduces no adverse system effects; thus, it is reasonable to focus on observable faults. 

For each $R_i$ with $n_i$ nodes, we assume at most $f_i$ nodes are faulty, where $n_i \geq 2f_i + 1$, and at most $F$ nodes are faulty in the whole system. 
Although $2f_i+1$ nodes are needed for recovery from $f_i$ faulty nodes, we only require $f_i+1$ \emph{replicas} per task. 
We follow standard security assumptions: 
each node $N_x$ has a public/private keypair $(\pi_x, \sigma_x)$, each node knows all other nodes' public keys, and there is a cryptographic hash function $H$.
We focus on attacks on compute nodes and assume that sensors, actuators, and network devices are correct. 

\subsection{Goals}  \label{subsec:goals}
\bsection{Goal 1: Ensure safety} Given a recovery bound $\recoverybound$ that the system can tolerate, our first goal is to ensure that if a fault occurs at $t$, the system will be in a safe state by $t + \recoverybound$ if no additional fault occurs during that interval. Specifically, by $t + \recoverybound$, the system has either resumed correct behavior, in which all faulty nodes and links have been excluded from running tasks or transmitting data, or switched to safe mode.

An adversary may strategically compromise nodes in succession to maximize damage---e.g., attacking a new node just before the system recovers from a previous fault. However, since faulty behaviors during recovery are also detected, the system is guaranteed to recover from any fault scenario within a maximum time of at most $F\cdot \recoverybound$. This BTR property aligns with prior work in fully synchronous settings~\cite{gandhi2021rebound,Gandhi2020BoundedtimeRF}.

\bsection{Goal 2: Maximize operational robustness} 
As discussed earlier, to maintain safety under poor inter-region network conditions, a region should switch to the safe mode if any task $\tau$ within the region does not receive any inter-region message within a timeout $\thbtimeout^{\tau}$. An attacker could exploit this mechanism by intentionally delaying or dropping messages to trigger unnecessary transitions to the safe mode even when the network conditions are normal. To maintain operational robustness, our second goal is to bound the number of inter-region messages a node can delay or drop within a given period, thereby limiting the adversary's ability to induce unwarranted transitions to safe mode.

\bsection{Solution} Since intra-region networks are synchronous and reliable, we can employ an existing synchronous BTR solution, such as \rebound~\cite{gandhi2021rebound}, to detect and recover from intra-region faults within each region. Our core contributions are protocols to handle (i) inter-region faults, and (ii) intra-region faults whose recovery requires coordination between different regions.
We first briefly describe \rebound (see~\cite{gandhi2021rebound} for more details), then give an overview of our proposed protocols.

\subsection{BTR for intra-region faults: Background}\label{sec:intra-btr}

\noindent\textbf{Replicas and multi-mode schedules. }
At the high level, each region is designed to operate in multiple (normal) modes, which specify how tasks are scheduled on the (correct) nodes under different fault scenarios. Each scenario represents the set of nodes $\mathbb{FN}$ and links $\mathbb{FL}$ that are known to have failed. Each task has $f_i + 1$ replicas to be scheduled on $f_i + 1$ different nodes in the same region. 
At runtime, a protocol is executed to detect faults when nodes start to deviate from their expected behavior. 
Whenever a new fault is detected, we perform recovery actions (e.g., migrate tasks, drop non-critical tasks, reroute traffic flows, switch to a new schedule) to switch the system to the appropriate new mode. 
These modes, computed offline, ensure that all application and protocol tasks meet deadlines and that communication latency constraints are satisfied.
Further, multi-mode real-time analysis is used to ensure that the mode transition finishes within the recovery bound $\recoverybound$. 

\bsection{Detection} 
For commission faults, all signed results of a replica task are forwarded to the 
sender's peer replicas. Since there are $f_i+1$ replicas and at most $f_i$ faulty nodes within $R_i$, at least one replica must be correct and can detect an output mismatch. That node will generate and disseminate a PoM ({\em proof of misbehavior}) message against the node that outputs the different result $r'$. Other correct nodes, on receiving the PoM, will replay the computation and get the replayed result $r$. If $r \neq r'$, they will add the node that outputs $r'$ to the set of faulty nodes $\mathbb{FN}$ that they are aware; otherwise, they add the node that generates the PoM to $\mathbb{FN}$ and disseminate it. 

For omission faults, since each node knows the schedule on every node and the maximum (intra-region)  network latency, if a message it expects is not delivered in time, the node declares an omission fault against the link between itself and the sender. The node will broadcast a PoM, and all nodes will add this link to their sets of faulty links $\mathbb{FL}$.

\bsection{Recovery} 
Each node stores the computed modes for all possible fault scenarios $(\mathbb{FN}, \mathbb{FL})$ locally. Each mode contains a task schedule  (e.g., what task replicas need to run on each node) and information on when it should expect messages from upstream nodes or replicas of its tasks. Upon a detected fault, each node switches to the new mode corresponding to the current fault scenario $(\mathbb{FN}, \mathbb{FL})$. 
If a fault occurs during recovery, say at $t' \in (t, t+\recoverybound]$, it will be detected and the protocol ensures the system recovers by $t' + \recoverybound$ (i.e., the time available for recovery is reset to $\recoverybound$ upon a new fault).

\bsection{Limitations} The above method works well within a region because the intra-region network is fully synchronous and reliable. However, both detection and recovery will fail when messages are exchanged across regions, as the inter-region networks may delay or drop messages---including both task outputs and PoM messages required for recovery.

\subsection{\system overview} \label{subsec:geoshield-overview}
\begin{figure}

    \centering
    \includegraphics[width=.78\linewidth]{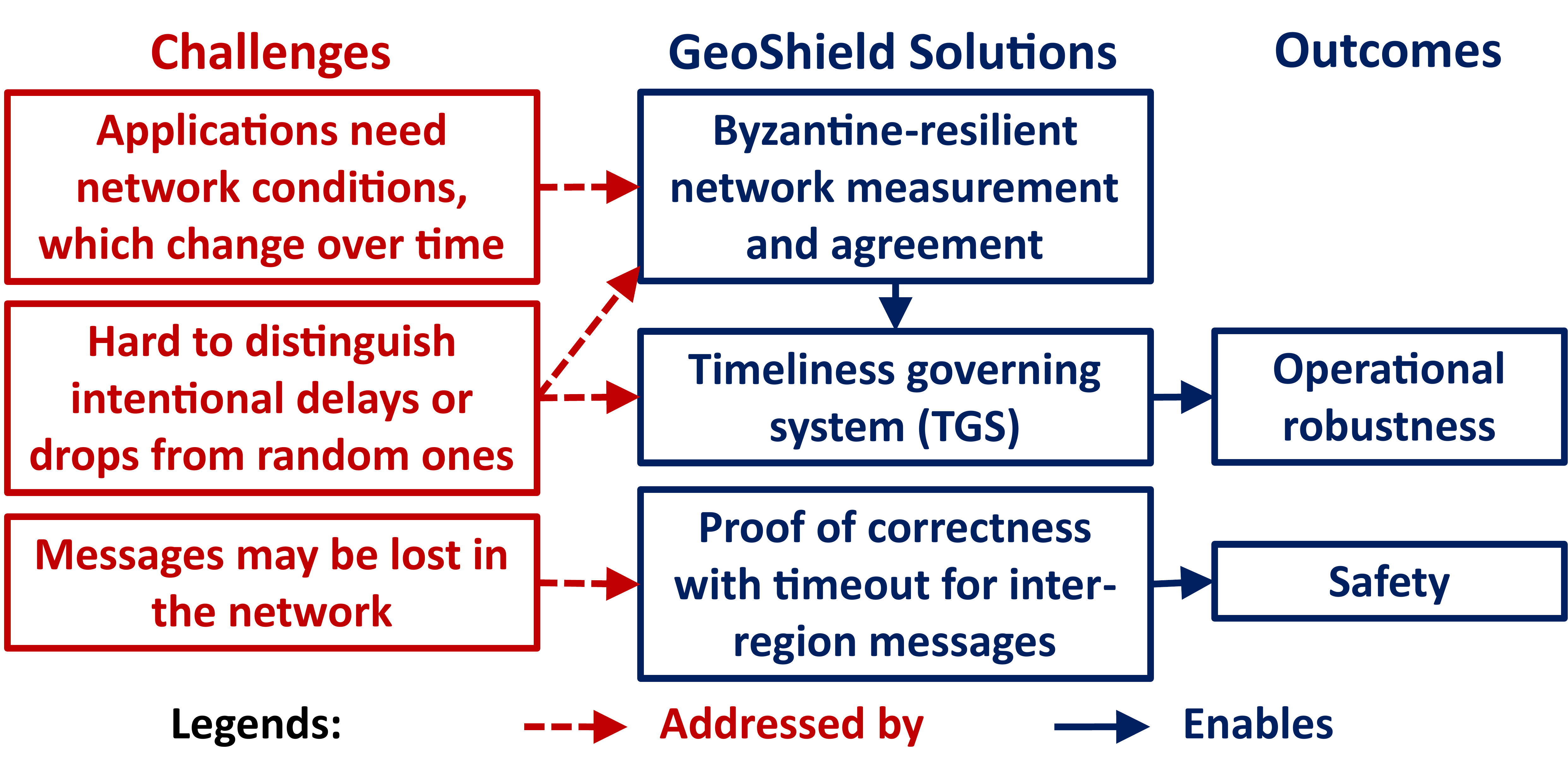}
 
    \caption{Challenges and solutions in \system. }
    \label{fig:challenge-sol-overview}

\end{figure}

\noindent Fig.~\ref{fig:challenge-sol-overview} highlights the core components of \system and how they achieve our safety and robustness goals.

\noindent\textbf{Byzantine-resilient network measurement.}
\system introduces a protocol for estimating the current probabilistic maximum inter-region network latency in the presence of Byzantine nodes. It guarantees that all correct nodes agree on the estimated value, and that the value closely approximates the actual one with high probability, despite disruptions by faulty nodes. A consistent and accurate latency bound estimate enables consistent and appropriate decision-making among correct nodes, which is essential for both normal CPS  operation~\cite{hu2021lars,ateya2019energy}  
and effective fault detection and recovery. 

\bsection{Inter-region omission fault detection}
Based on the agreed network latency value, nodes can identify inter-region messages that take significantly longer than expected to arrive. The senders and receivers of such messages are marked as \emph{suspicious} for omission faults. \system introduces a distributed {\em timeliness governing system} (TGS) 
to track nodes' behaviors and to bound the number of suspicious behaviors over both short and long time frames. TGS limits the ability of compromised nodes to trigger unnecessary transitions to safe mode, thereby maintaining operational robustness.

\bsection{Inter-region commission fault detection}
Unlike intra-region networks where fault evidence (c.f. Sec.~\ref{sec:intra-btr}) can be reliably disseminated, inter-region evidence messages may be lost or delayed due to unreliable networks. Since evidence messages are aperiodic---sent only when a fault is detected---their loss or delay means that nodes in the receiving regions may remain unaware of the fault past the recovery time bound, causing safety violations. To address this, we introduce the concept of {\em proof of correctness} (PoC), a cryptographic endorsement attached to periodic heartbeats. 
Receivers of inter-region messages can validate their correctness by verifying the timely arrival of their corresponding PoCs. Missing or delayed PoCs indicate faults, which trigger timely recovery.

\bsection{Bounded-time recovery}  
Finally, we design mechanisms for nodes in each region to perform recovery in response to detected faults in the same region or in a remote one. We formally show that our detection and recovery protocols ensure that the system can recover within bounded time.

\section{Byzantine-Resilient Network Measurement}\label{sec:measurement}

\subsection{Probabilistic bounded jitter property} \label{sec:jitter-property}

\noindent In safety-critical CPS, operators often deploy dedicated networks under their full control~\cite{nokia-backbone,hollysys-brochure}. This enables them to schedule and analyze workloads~\cite{DBLP:journals/access/MamaneFGBBM22,zhou2020survey} to ensure predictable latency. Further, as routing in such networks is typically stable, a major source of network jitter is eliminated~\cite{pucha2007understanding,reda2020path}. Thus, the primary remaining sources of delay variance are transient or permanent hardware faults, node mobility, and random signal interference. 
Notably, we observe that unless they occur during or between the transmissions, these sources will not cause significant variance in the delays of two packets that are sent closely together between the same pair of regions.
This observation aligns with findings from existing measurement studies in geo-distributed systems, which show that the jitter of most packets~\cite{becker2014measurement,huang2012close,torres2020qos,reda2020path}, particularly successive ones~\cite{torres2020qos}, is typically small. This assumption is also commonly used in network link capacity estimation~\cite{dovrolis2004packet}.


\begin{assumption}[Probabilistic bounded jitter]\label{assume:jitter}
    For any two inter-region messages, $m$ from $N_x$ to $N_y$ and $m'$ from $N_{x'}$ to $N_{y'}$, the difference in their network delay is less than $\Dinter$ with probability $\ge$ 
    $\pnorm$ if: \circled{1} the senders $N_x$ and $N_{x'}$ are in the same region, and receivers $N_y$ and $N_{y'}$ are in the same region, and  \circled{2} the two messages are sent within $\epsilon_t$ time units from each other, where $\epsilon_t$ is some small value.
\end{assumption}


\vspace{.25ex}
\noindent{\bf Experimental validation.} 
As we do not have access to a real geo-distributed CPS network infrastructure, we set up servers on Cloudlab~\cite{duplyakin2019design} (in MA, SC, UT, and WI), Google Cloud (in IA and OR), and Microsoft Azure (in VA and WA). 
For each pair of regions, we sent messages between them every second. 
In our experiments, $\epsilon_t$ could range from 100~$\mu s$ to 1 ms, due to clock skew and the processing time variance.

\begin{table}[t]
\centering \footnotesize
\begin{adjustbox}{scale=.87}
\begin{tabular}{|c|c|c|c|c|c|}
\hline
\textbf{Route} & \textbf{Median} & \textbf{99.9 Perc.} & \textbf{Route} & \textbf{Median} & \textbf{99.9 Perc.} \\ \hline
\textbf{MA $\to$ IA} & 238 & 848  & \textbf{SC $\to$ UT} & 15  & 133  \\ \hline
\textbf{UT $\to$ IA} & 337 & 1314 & \textbf{WI $\to$ UT} & 16  & 137  \\ \hline
\textbf{WI $\to$ IA} & 106 & 896  & \textbf{MA $\to$ VA} & 972 & 5184 \\ \hline
\textbf{SC $\to$ MA} & 14  & 264  & \textbf{WI $\to$ VA} & 7   & 5021 \\ \hline
\textbf{UT $\to$ MA} & 18  & 263  & \textbf{MA $\to$ WA} & 558 & 4672 \\ \hline
\textbf{WI $\to$ MA} & 22  & 220  & \textbf{UT $\to$ WA} & 12  & 2406 \\ \hline
\textbf{MA $\to$ OR} & 277 & 999  & \textbf{MA $\to$ WI} & 65  & 283  \\ \hline
\textbf{WI $\to$ OR} & 15  & 709  & \textbf{SC $\to$ WI} & 28  & 211  \\ \hline
\textbf{MA $\to$ UT} & 22  & 328  & \textbf{UT $\to$ WI} & 36  & 218  \\ \hline
\end{tabular}
\end{adjustbox}
\caption{The latency difference (in $\mu$s) of successive packets between regions measured over one day.}\label{tbl:jitter}
\end{table}


\noindent{\em Results.} Table~\ref{tbl:jitter} provides latency difference statistics for successive packets transmitted between different regions over a 24-hour period, where the latency between each pair of nodes is measured once per second.\footnote{The nodes hosted on Google Cloud and Microsoft Azure are virtual machines, which exhibit higher latency variance due to virtualization overhead. Results from regions with highly unstable routing are excluded, as real-world real-time CPS typically have dedicated networks with significantly more stable routing than the public Internet.} 
Results show that the latency difference between successive packets is small with high probability. Detailed latency graphs for an example trace that further illustrate this property are available in Appendix~\ref{app:jitter-meas}.

In a practical deployment, operators should plot the CDF of the latency difference profiled (e.g., Fig.~\ref{fig:latency}) and select the pair $(\pnorm, \Dinter)$ from the CDF---with $\pnorm$ being the target percentile and $\Dinter$ the corresponding latency difference bound---that ensures safe and robust operation (see Sec.~\ref{sec:discussion}).

\subsection{Measurement protocol: Basic ideas} 
\noindent At the high level, the network measurement task in each region $R_i$ is done by a set of $f_i+1$ nodes, called \textit{measurers}. 
Each measurer periodically sends heartbeat messages to measurers in external regions that require communication with $R_i$. 
The heartbeats are scheduled to be sent in rounds, at a predetermined time $\tsendn$ for round $n$, which are known to all nodes. 
Upon receiving a heartbeat from another region $R_j$ for a round $n$, the receiver in $R_i$ estimates the network latency from $R_j$ to $R_i$ by taking the difference between the heartbeat arrival time and $\tsendn$, 
and sends its estimated value to other nodes in $R_i$. 
Each node decides on a value based on the received estimates.



\begin{figure*}[t]
    \centering
        \begin{minipage}{0.7\linewidth}
        \hspace{-2ex}
            \includegraphics[width=\linewidth]{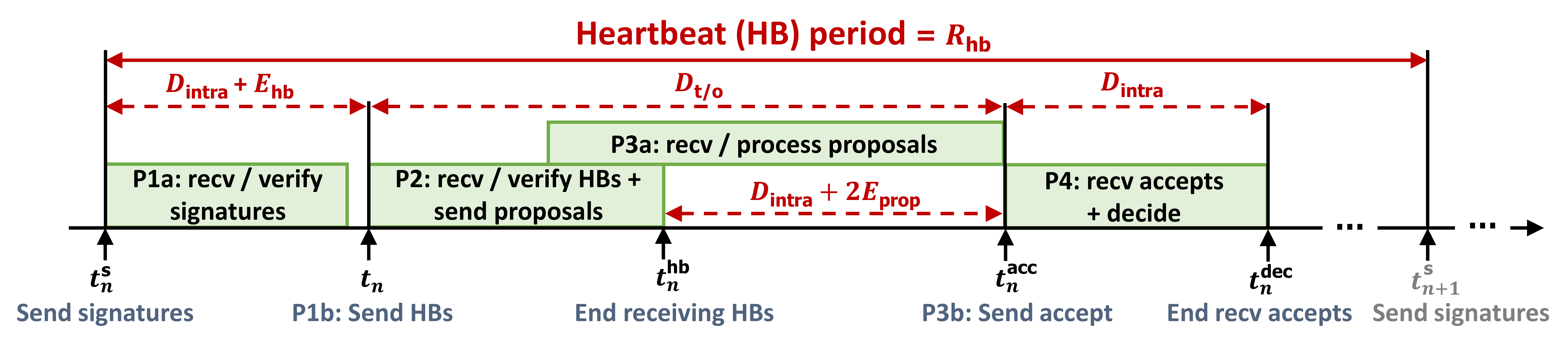}
            \caption{Different phases of the measurement protocol.}
            \label{fig:timeline}
        \end{minipage}
    \begin{minipage}{0.25\linewidth}
    \centering
    \raisebox{-9.5ex}{
        \scalebox{0.72}{
            \begin{tabular}{|l|l|}
            \hline
            \textbf{$\pnorm$}      & The probability in Assumption~\ref{assume:jitter}  \\ \hline
            \textbf{$\Dinter$}     & The jitter in Assumption~\ref{assume:jitter}       \\ \hline
            \textbf{$\tintra$}     & Max intra-region latency                  \\ \hline
            \textbf{$\tinterval$}  & Period of sending HBs               \\ \hline
            \textbf{$\tsendn$}     & Time to send HBs (from round $n$)          \\ \hline
            \textbf{$\thbtimeout$} & Timeout to accept latency since $\tsendn$ \\ \hline
            \textbf{$\tdecide$}    & The decided inter-region latency                       \\ \hline
            \end{tabular}
        }
    }
    \vspace{.75ex} 
    \captionof{table}{Notations.}\label{tbl:notations}
    \end{minipage}
\end{figure*}

\bsection{Challenge and solution}
Achieving consensus and accuracy on the latency estimate is non-trivial. A faulty node may send a heartbeat much earlier or much later than its scheduled time to cause the receiver to under-estimate or over-estimate the latency. Likewise, it may send different estimates to different nodes to create inconsistency.
To prevent under-estimation, we require that a valid heartbeat message from a region $R_i$ in round $n$ must include valid signatures on $n$ from $f_i + 1$ measurers in $R_i$, and that these signatures must be exchanged among measurers at a specific time $t_{n}^s$ shortly before the heartbeat scheduled time $t_n$. This guarantees that a faulty sender cannot generate a valid heartbeat much earlier than $t_n$. 
We also require the heartbeat receivers to forward the heartbeat immediately after receiving it, so that other nodes can verify whether their claimed estimated latency values are reasonable. To prevent over-estimation, any pair of sending and receiving measurers whose heartbeat experiences a much higher latency than those between other pairs of measurers (which is unlikely due to the probabilistic jitter assumption) will be suspected as faulty. Finally, to circumvent commission faults, an estimated value is only chosen if all measurers agree on it.

Fig.~\ref{fig:timeline} shows the timeline in each round of the protocol,   
and Table~\ref{tbl:notations} lists key notations. We first state our assumptions, which typically hold for CPS (see Sec.~\ref{subsec:sys-model}). 

\bsection{Assumptions}
We assume that the maximum clock skew among all nodes is $\Dsyn$. 
The intra-region message latency is in the range of $[\tintra - \Dintra, \tintra]$.
The time each node takes to verify signatures and construct its heartbeat message is in the range of $[\thb-\Dhb, \thb]$. 
The total time each node takes to process the received heartbeats and send proposals, or to verify the received proposals and send accept messages, is in the range of $[\tprop - \Dprop, \tprop]$. 

\bsection{Notations} $\tinterval$ and $\thbtimeout$ denote the heartbeat period and heartbeat timeout, respectively; they should be chosen 
such that the recovery bound is met (see Sec.~\ref{sec:discussion}). 
The following denote the predetermined {\em scheduled} time of different actions during each measurement round $n$:   
$\tsendn$ is time to send the heartbeat, i.e., $\tsendn = n \cdot \tinterval + t_0$ where $t_0$ is the time to send the first heartbeat. 
$t_{n}^s = \tsendn - \tintra - \thb$ is time to exchange the signature on $n$.  
$t_n^{\mathsf{hb}}$ is time to stop both receiving heartbeats and sending {\em proposal} messages.
$t_n^{\mathsf{acc}} = \tsendn + \thbtimeout$ is time to send the {\em accept} message. 
$t_n^{\mathsf{dec}}$ is the time to stop receiving accept messages and the latest time 
to decide on an estimated latency bound.

\subsection{Details of the measurement protocol} \label{subsec:meas-phases}
\noindent In each round $n$, each measurer $N_x$ in a region $R_i$ executes through several phases:

\vspace{.25ex}
\noindent{\bf Phase 1a: Signature exchange.} At time $t_n^s$, $N_x$ multicasts its signature $\sigma_x(n)$ to all peer measurers, then waits for signatures from other measurers. 
Whenever it receives a signature from another measurer, $N_x$ adds the received signature and the sender into the signature set $S_n$ for this round.   
At time $t_n^s + \tintra$, if $N_x$ has not received a signature from some peer measurer, it declares 
omission fault against that measurer and triggers intra-region recovery. Otherwise, it verifies the signatures in $S_n$.
If a signature from $N_y$ is an invalid signature on $n$, $N_x$ declares an omission fault
against the link between $N_x$ and $N_y$, disseminates its evidence, and performs intra-region recovery. 
If $S_n$ contains $f_i+1$ valid signatures on $n$, $N_x$ signs a hash of $S_n$ and constructs its heartbeat containing its ID $x$, the set of signatures
$S_n$, and the signed hash (i.e., $\bigl\langle x, S_n, \sigma_x( H(x,S_n))\bigr\rangle$). 

\vspace{.25ex}
\noindent{\bf Phase 1b: Send heartbeat.} At time $t_n$, $N_x$ multicasts its constructed heartbeat 
to all measurers in all downstream regions that $R_i$ communicates with. 
$N_x$ will then wait for heartbeats from each upstream region $R_j$.

\vspace{.25ex}
\noindent{\bf Phase 2: Receive heartbeats and send proposals.}
When $N_x$ receives a heartbeat $m$ from a measurer $N_y$ in $R_j$, 
it logs $m$'s arrival time as $t_{\mathsf{recv}}$ and verifies $m$. If $m$ is invalid (i.e., not signed by $N_y$ or does not contain $f_j+1$ valid signatures on $n$), $N_x$ declares an inter-region commission fault against $N_y$ and triggers fault recovery. 
Otherwise, 
$N_x$ proposes $d = t_{\mathsf{recv}} -\tsendn$ as the latency from $R_j$ to $R_i$. It 
constructs a signed {\em proposal} containing its node ID $x$, the proposed latency $d$, and the received heartbeat $m$. $N_x$ then sends the signed proposal $\bigl\langle x,d, m, \sigma_{x}(H(x,d,m))\bigr\rangle$ to all peer measurers and to $f_i$ other nodes (named \emph{log keepers}) in its region $R_i$. (The log keepers log the messages and 
only process them in case of disputes among measurers.) 
Additionally, $N_x$ logs its proposal and records the smallest latency $P_{\mathsf{min}}$ it has proposed in this round.

Each measurer processes heartbeats for round $n$ up to time $t_{n}^{\mathsf{hb}}$ only; this is to leave sufficient time for other measurers to process latency proposals by $t_{n}^{\mathsf{acc}}$, the scheduled time for sending the accept message.

\vspace{.25ex}
\noindent{\bf Phase 3a: Receive and process proposals.}
Whenever $N_x$ receives a proposal from a peer measurer $N_{x'}$, it verifies the signatures on the proposal and on the included heartbeat $m$. If the signatures are valid, $N_x$ checks if the proposed latency 
is reasonable. Specifically, because (1) the time $t$ at which $N_x$ receives the proposal is at most $\tprop + \tintra$ time units 
after $N_{x'}$ receives $m$, and (2) $m$ is sent from the upstream region at $t_n$, only proposed latency values no smaller than $t - \tsendn - \tprop - \tintra$ are reasonable. Taking the maximum clock difference $\Dsyn$ into account, $N_x$ will deem proposed values that are at least 
$d_{\min} = t - \tsendn - \tprop - \tintra - \Dsyn$ reasonable.
If the proposed latency is reasonable, $N_x$ logs the proposal. Besides, $N_x$ also keeps track of the smallest reasonable proposed latency $A_{\mathsf{min}}$ in this round.

\vspace{.25ex}
\noindent{\bf Phase 3b: Send accept.}
At time $t_{n}^{\mathsf{acc}}$, $N_x$ calculates an estimate of the probabilistic worst-case network latency 
between $R_j$ and $R_i$:  
$\taccept = \min(A_{\mathsf{min}}, P_{\mathsf{min}}) + \Dinter$.
It then prepares and broadcasts a signed {\em accept} message $\bigl\langle x,n, \taccept, \sigma_x(H(x,n,\taccept))\bigr\rangle$ 
to all nodes in its region. If no proposals have been sent or received, $\taccept$ will be a special indicator for timeout.

\vspace{.25ex}
\bsection{Phase 4: Decide}
Once a node in region $R_i$ receives $f_i+1$ accept messages with valid signatures for the same accepted latency 
$\taccept$ in round $n$, the node will decide on this latency $\tdecide = \taccept$
as the current probabilistic worst-case inter-region network latency. 
If it does not receive an accept message from a measurer by $t_{n}^{\mathsf{dec}}$, it declares an omission fault against the measurer. 
If a measurer receives a different value of $\taccept$ from its own, it declares a fault and recovers from it. (See details in Theorem~\ref{thm:agreement} and Appendix~\ref{subsec:app-detection-recovery-meas}.)

We next state key properties and guarantees of the protocol.

\bsection{Properties}
Lemma~\ref{lem:no-early} bounds the amount of time a node can send a valid heartbeat earlier than scheduled, as a direct result of the signature exchange. Theorem~\ref{thm:meas-accuracy} bounds the difference between a decided latency estimate and actual latency. Finally, Theorem~\ref{thm:agreement} establishes agreement on the decided latency. Due to space constraints, we provide the detailed proofs and descriptions of the detection and recovery protocols used during measurement in 
Appendix~\ref{app-sec:meas}.

\begin{lemma}[Early heartbeat] \label{lem:no-early}
     It is impossible for a node to send a valid heartbeat message more than $\tearly = \Dsyn + \Dintra + \Dhb$  time units before the scheduled time. 
\end{lemma}

\begin{theorem}[Accuracy] \label{thm:meas-accuracy}
    Suppose the actual probabilistic maximum inter-region latency in round $n$ is $D_{\mathsf{real,n}}$, then
    
    $\tdecide \geq D_{\mathsf{real,n}} -  (\tearly + \Dprop + \Dintra + \Dsyn)$; and 
    
    $\tdecide \leq D_{\mathsf{real,n}}+ (\Dinter+\Dsyn) $, with probability $\ge \pnorm$.

\end{theorem}
\begin{theorem}[Consensus on latency] \label{thm:agreement}
    The following properties hold: 
    1) At $t_{n}^{\mathsf{dec}}$, either all correct measurers agree on the same probabilistic maximum inter-region latency $\tdecide$, or at least one faulty node or faulty link will be detected; and 2) after the fault is detected, a new latency can be decided within bounded time, all correct measurers and log keepers will decide on the same latency, and the decided value is from a valid proposal.
\end{theorem}

\section{Handling Inter-region Faults}

\begin{figure*}
    \begin{minipage}{1\linewidth}
        \centering
        \includegraphics[width=0.7\linewidth]{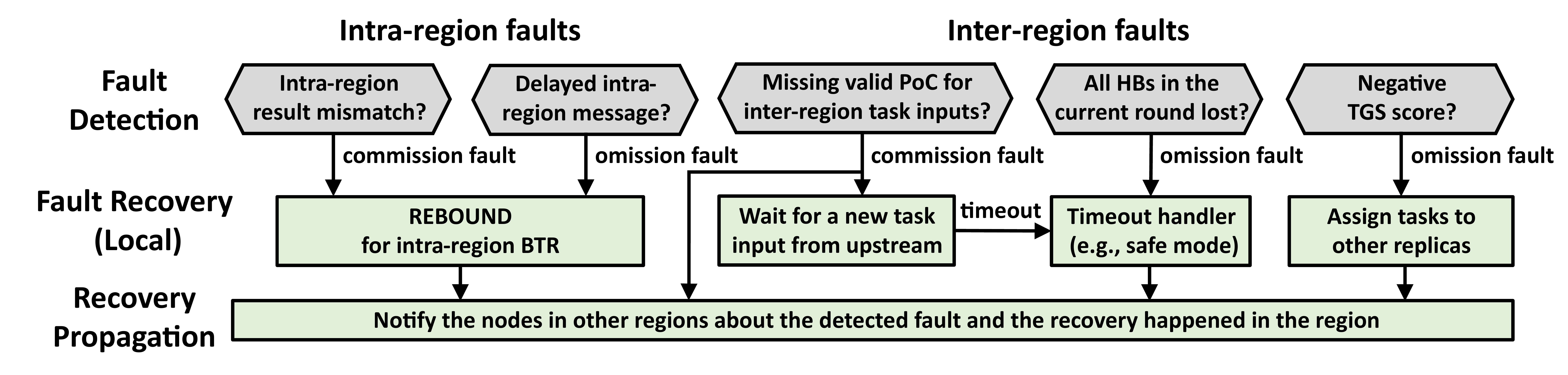}
        \caption{Mechanisms to detect and recover from different types of faults.}
        \label{fig:overview-det-rec}
    \end{minipage}
\end{figure*} 

\begin{figure}
        \centering
         \includegraphics[width=\linewidth]{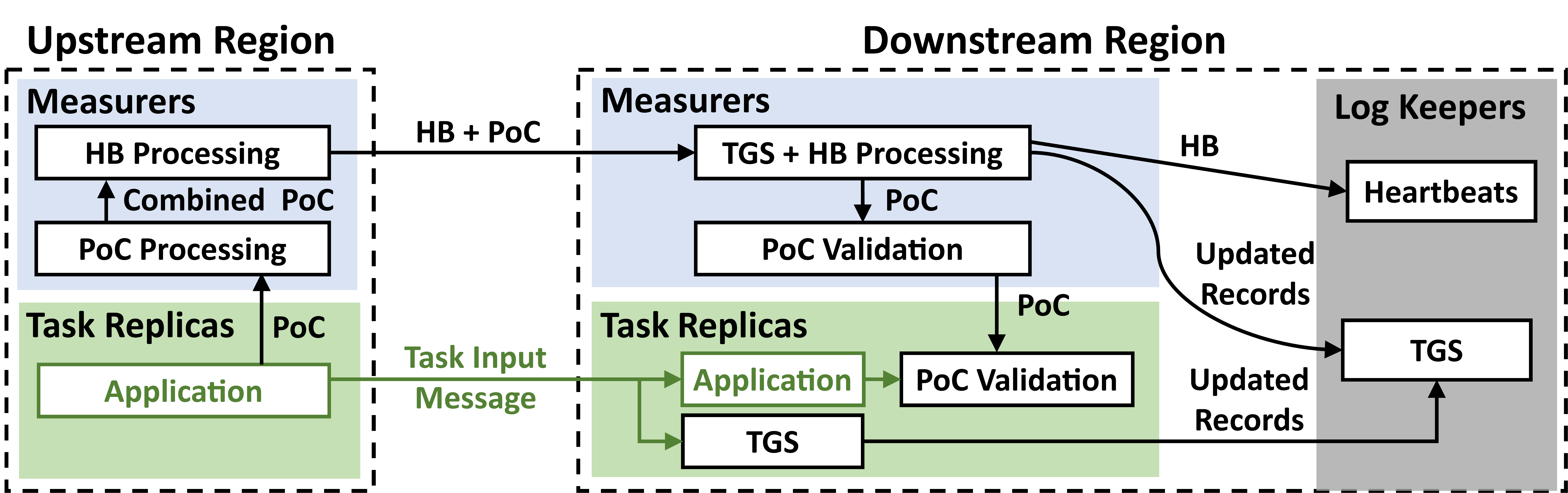}
        \caption{Inter-region task dataflow.}\label{fig:detection-flow}
\end{figure}

\noindent 
Fig.~\ref{fig:overview-det-rec} and~\ref{fig:detection-flow} give an overview of how different types of faults are handled in \system. As discussed in Sec.~\ref{subsec:goals}--\ref{subsec:geoshield-overview}, \system leverages \rebound for detecting and recovering from intra-region faults, and it uses two newly proposed mechanisms---PoC and TGS---for handling inter-region faults. The first, PoC, provides a way for nodes to verify the correctness of inter-region messages, which is necessary for commission fault detection and safety guarantees. The second, TGS, leverages the probabilistic maximum network latency estimates to bound the number of omission faults committed by a node, which is necessary to minimize unwarranted transitions to safe mode, thereby enhancing operational robustness. 

\subsection{Handling commission faults}\label{subsec:inter-region-comm}
\bsection{Introducing proof of correctness (PoC)}
Whenever a node $N_x$ in region $R_i$ sends a message $m$ containing the input for a downstream task $\tau'$, it also sends a PoC for $m$ to $f_i+1$ measurers in $R_i$.
(Note that $m$ should include a unique identifier---such as the IDs of the task and job that produce $m$ as output---to prevent faulty nodes from reusing old PoCs.) A PoC for $m$ is a tuple $(\tau', H(m), \Sigma)$, where $\tau'$ is a downstream task ID,  $H(m)$ is a hash of $m$, and $\Sigma$ is a set of signatures (on the message hash) generated by nodes that endorse 
$H(m)$.

\vspace{.25ex}
\bsection{Fault detection} 
We now describe the dataflow related to PoCs, shown in Fig.~\ref{fig:detection-flow}. For accountability, message senders always sign their messages. Consider a task $\tau$ in region $R_i$ that produces output to a downstream task $\tau'$ in region $R_j$. Let $N_x$ and $N_y$ be replicas of $\tau$ and $\tau'$, respectively.

{\em Upstream task replicas.} Whenever $N_x$ completes a job of $\tau$, it sends the output $m$ (i.e., an input for $\tau'$) to all replicas of $\tau'$. Additionally, $N_x$ constructs a PoC for $m$---$(\tau, H(m), \{ \sigma_x(H(m)) \})$---and sends it to all measurers in $R_i$. Failing to do so in time (discussed later) will result in an intra-region omission fault against $N_x$.

{\em Upstream measurers.} Upon receiving the PoCs for $m$ from $\tau$'s replicas, each measurer checks if the PoCs have the same hash $H(m)$ and valid signatures on $H(m)$. If not, it declares an intra-region commission fault; otherwise, it adds all the $f_i$ signatures received, plus its own, to the set $\Sigma$, resulting in a new PoC with $f_i+1$ valid signatures (called final PoC), which it will include as part of the heartbeat message to region $R_j$ in round $n^{\star}$, where $n^{\star}$ is determined as follows. 

Let $t_m$ be the deadline for $N_x$ to send $m$ to the downstream task $\tau'$, $E_{\mathsf{poc}}$ be the maximum time to generate a PoC, and $E_{\mathsf{sig}}$ be the maximum time to verify the PoC and generate a signature for signature exchange. Then, the round $n^{\star}$ of the heartbeat containing $m$'s final PoC is defined as the smallest value $n$ that satisfies 
$\tsendn \geq t_{m} + \tpocgap$, where $\tpocgap = E_{\mathsf{poc}} + 2\tintra + E_{\mathsf{sig}}+\thb$.
This condition ensures that all measurers agree on the same round $n^{\star}$ for the heartbeat containing $m$'s PoC, and that they have sufficient time to verify the PoCs and generate the signatures to exchange. 

To ensure PoCs are included in the heartbeats correctly,
a measurer $N_x$ should perform a slightly modified signature exchange phase of the latency measurement protocol (Phase 1a)---it exchanges and verifies the signatures of $\sigma_x(n^{\star},\text{PoCs})$ instead of simply  $\sigma_x(n^{\star})$, where PoCs is the concatenation of all final PoCs (ordered by task ids) to be included in the heartbeat in round $n^{\star}$. At $t_{n^*}^s$, $N_x$ sends the heartbeats, with PoCs attached, to downstream measurers.

{\em Downstream measurers.} 
When a measurer in $R_j$ receives a PoC for an input of $\tau'$ from a measurer in $R_i$, it checks if it is a valid final PoC by checking if $f_i+1$ valid signatures exist.
If so, it forwards the PoC to the replicas of $\tau'$; otherwise, it declares a commission fault against the sender. 
For all subsequent PoCs for $\tau'$, 
it checks consistency with the first validated one and declares a commission fault against any sender of an inconsistent PoC.

{\em Downstream task replicas.} 
Each downstream replica $N_y$ maintains a set $S_m$ of messages containing the next input to $\tau'$. When it receives $m$ for $\tau'$ from an upstream replica, if $S_m$ is empty, it uses $m$ as the input to $\tau'$. If $m\notin S_m$, it adds $m$ to $S_m$ and forwards $m$ to all replicas of $\tau'$. 

When $N_y$ receives a PoC, it first checks if it is a valid final PoC; if not, it declares an intra-region commission fault against the measurer that forwarded the PoC. Next, for each $m \in S_m$, $N_y$ checks if its hash matches the hash in the (valid) PoC. If not, $N_y$ declares an inter-region commission fault against $m$'s sender. The first {\em valid} PoC that $N_y$ receives will be forwarded to all replicas of $\tau'$, in case a faulty measurer selectively forwards the PoC to only a subset of the replicas. 

Let $n$ be the round of the heartbeat expected to contain the PoC for the input of $\tau'$. At  $t_{n}^{\mathsf{dec}}$, if $N_y$ has not received a valid final PoC for some message $m \in S_m$, and if the decided latency $\tdecide$ of $R_i$ to $R_j$ does not indicate a timeout, $N_y$ will mark $m$ as incorrect and declares a fault against its sender in $R_i$.

\vspace{.25ex}
\bsection{Fault recovery}
When a replica $N_y$ of a task $\tau'$ detects an incorrect input message for $\tau'$ from region $R_i$, $N_y$ starts a recovery propagation process (c.f. Sec.~\ref{subsec:btr}) to inform nodes in $R_i$ about the fault. If all messages in $S_m$ are incorrect (i.e., without valid PoCs), $N_y$ also adds a request for a new correct input for $\tau'$ in the recovery propagation request it sends (otherwise, it uses the correct message for $\tau'$). 

If requested, $\tau$'s replicas in $R_i$ should send a new correct input---with $f_i+1$ signatures endorsing it---to the replicas of $\tau'$.  
If a new correct input arrives at $N_y$ before a pre-configured timeout $\thbtimeout^{\tau'}$, $N_y$ will forward the input together with its endorsements to all replicas of $\tau'$ and use it for $\tau'$. Otherwise, $N_y$ will trigger a safe mode transition. 
In addition, a safe mode will be triggered if the decided latency $\tdecide$ is the timeout indicator due to bad network conditions, i.e., no heartbeat (thus, no PoC) arrives in that round.

\vspace{.25ex}
\bsection{Properties} 
As stated in Theorem~\ref{thm:consensus-poc}, \system ensures that all inter-region commission faults will be detected, and all correct nodes agree on the same state that the system should be in. A proof is available in Appendix~\ref{sec:appendix-consensus-poc}.

    

\begin{theorem}[Consensus on correctness] \label{thm:consensus-poc}
     All correct nodes in a downstream region $R_j$ agree on 1) whether an input message $m$ from an upstream region $R_i$ is correct, and 2) whether a safe mode should be triggered. 
\end{theorem}

\subsection{Handling omission faults with TGS}
\label{subsec:reputation}

\bsection{TGS design overview}
The goal of TGS is to detect omission faults in nodes that handle inter-region (measurement and application) tasks and reassign these tasks from faulty nodes to others, to prevent faulty nodes from triggering unwarranted safe mode transitions. Towards this, TGS introduces two key metrics: a \emph{flag counter} and a \emph{timeliness score}. The flag counter of a node tracks how many times it has been flagged for frequent inter-region message loss or delay. This counter is used in task reassignment, and nodes with lower counters are prioritized for running inter-region tasks. 
Each node is also assigned a timeliness score for each inter-region task it runs, which reflects how well the node has behaved since its most recent assignment with the task  (higher means better).

To maximize message delivery timeliness, TGS ensures that inter-region task replicas have good records, i.e., low flag counters and high timeliness scores.
(Note that measurers are replicas of the measurement task.) 
Within each region $R_i$, each inter-region task replica runs a local copy of TGS to track the behaviors of both its peer replicas in $R_i$ and replicas of upstream tasks in upstream regions. Additionally, another $f_i$ nodes are selected as log keepers that log the updated records. If a dispute arises, the log keepers and the $f_i+1$ task replicas will form a correct majority that can verify the logged record updates and identify faulty nodes with incorrect records. 

\vspace{.25ex}
\bsection{Detecting faults with timeliness scores}
Each node has an initial (and maximum) score $s_{\mathsf{init}} = 1$ for each inter-region task $\tau$ it is assigned. When the score is no longer positive, the node will be \emph{flagged}, and the task will be assigned to a new node. 
We define the behavior of nodes as follows: 
Suppose a node $N_s$ running $\tau$ in region $R_i$ needs to send a message $m$ to another node $N_r$ running $\tau'$ in a downstream region $R_j$ at $t$. 
The behavior of $(N_s,N_r)$ w.r.t. $(\tau, \tau')$ is \emph{suspicious} if $N_r$ claims that it does not receive $m$ by $t + \tdecide$, where $\tdecide$ is the decided probabilistic maximum latency in the most recent round of measurement; 
otherwise, the behavior 
is \emph{normal}.

At $t+\tdecide$ (or $t+\thbtimeout$ if $\tau$ and $\tau'$ are measurement tasks), $N_r$ should send the claims including all node/task pairs $(N_s: \tau,N_r:\tau')$ with suspicious behaviors to other replicas plus $f_j$ log keepers (to ensure the majority of correct nodes receive the claims). 
When receiving such a claim, the replicas decrease the scores of $N_s$ for $\tau$ and of $N_r$ for $\tau'$ by $\spenaty$ in their TGS. For all other pairs that do not exist in any of the claims, the replicas increase both scores by $\saward$.
The log keepers store the claims when receiving them in case of score inconsistency later. 
We note that TGS behaviors are based on the message receivers' claims, since other nodes have no way to validate whether an inter-region message did or did not arrive at a node. Since it is impossible to distinguish whether $N_s$ or $N_r$ delays/drops $m$, we penalize both $N_s$ and $N_r$. A faulty $N_r$ may manage to decrease the score of a correct $N_s$, but we ensure that $N_r$ cannot do this indefinitely without itself being flagged.

We set an upper bound $s_{\max} = 1$ for the scores, so a faulty node cannot accumulate a high score by first being correct but later dropping many messages in a short period while maintaining a positive score.
To choose $\saward$ and $\spenaty$, we introduce parameters $\alpha \in (0,1], \beta \in \mathbb{N}^{+}$, and let 
$ \spenaty= s_{\max} / \beta$ and $\spenaty / \saward = \alpha \cdot \pnorm / (1-\pnorm) $.
Intuitively, $\beta$ restricts the number of consecutive suspicious behaviors, and $\alpha$ controls how much TGS punishes suspicious behaviors relative to awarding normal ones (larger $\alpha$ means more penalty).

\begin{theorem}[TGS Properties] \label{thm:TGS}
The following properties hold:
    \begin{enumerate}
        \item \emph{(Correct pairs expect to be awarded).}
        Let $E(N)$ be the expected score change of node $N$ in each
        round. For all pairs of correct nodes $(N_s, N_r)$, where $N_s$ is the sender and $N_r$ is the receiver of an inter-region message, $E(N_s) > 0$ and $E(N_r) > 0$. \label{item:good-score}
        \item \emph{(Long-term bound).} To keep a positive score, a node
        must behave normally for at least $p'= \frac{\alpha \cdot
        \pnorm}{1+(\alpha-1)\pnorm}$ of the messages in the long term. \label{item:long-term-bound}
        \item \emph{(Short-term bound).} To keep a positive score,
        within any time window of size $w = \beta + k + k\cdot \alpha \cdot \frac{\pnorm}{1-\pnorm}$, 
        a node must behave normally for more than $w-(\beta + k)$ messages ($\forall k \in \mathbb{N}^*$)\label{item:short-term-bound}
    \end{enumerate}
    
\end{theorem}
\noindent Theorem~\ref{thm:TGS}-\ref{item:good-score} guarantees that correct nodes will not be frequently flagged (to avoid unnecessary task reassignment).
Theorems~\ref{thm:TGS}-\ref{item:long-term-bound} and~\ref{thm:TGS}-\ref{item:short-term-bound} restrict the number of suspicious behaviors of a node in the {\em long term} and the {\em short term}, respectively. A proof of the theorem is in Appendix~\ref{sec:appendix-tgs-properties}.\\


\bsection{Recovery with task reassignment based on flag counters}
Recall that each node is assigned a flag counter that indicates how many times it has been flagged. Initially, the counter is zero, and it should be increased by one every time the node is flagged. As discussed above, each node $N_{x}$ that executes inter-region tasks periodically updates the timeliness scores for both its peer replicas and replicas of upstream tasks. Upon each update, if $N_{x}$ detects a node $N_y$ with a non-positive score for a task $\tau$, it performs recovery as follows: 

1) If $N_y$ is in the same region $R_i$ as $N_x$: Then, $N_{x}$ starts to prepare for a flag (and reassignment) proposal. It finds the node $N_z$ in its region with the smallest counter that has enough spare computation resources to run $\tau$, with node ID being the tiebreaker. It will then broadcast a signed message---containing a proposal to flag $N_y$ and to reassign $\tau$ to node $N_z$---to all nodes within its region. Since TGS runs on all replicas in the same way, all correct replicas will also broadcast the same proposal. When a node receives $f_i+1$ matching proposals, it starts migrating the tasks and updates the flag counter of the flagged node $N_y$ according to the proposal. If a node receives an incorrect proposal (commission fault) or does not receive a matching proposal in time (omission fault) from a certain replica, it declares a fault against that replica. Since there are $f_i$ log keepers and $f_i+1$ replicas for $\tau$, and at most $f_i$ nodes can be faulty, we have a majority of nodes with the correct TGS records. 
Moreover, as the reassignment strategy is deterministic, any node not sending the same correct proposal will be detected.

2) If $N_y$ is in an upstream region $R_j$: Then, $N_{x}$ will generate and sign a flag proposal including $N_y$ and $\tau$. It then starts a recovery propagation (i.e., the process to send fault information to other regions, see Sec.~\ref{subsec:btr}). The nodes in $R_j$, after verifying the recovery propagation, will reassign $\tau$ to another node $N_z$ using the above selection strategy. 
They will then start another recovery propagation to inform nodes in regions that exchange messages with $\tau$ about the reassignment. After verifying the propagation, these receivers will adjust the dataflow. The replicas of $\tau'$ will also initialize the score of $N_z: \tau$ to $s_{\mathsf{init}}$ in their TGS and start tracking its behavior.

\subsection{Recovery propagation}
\label{subsec:btr}

\noindent Unlike single-region systems, \system requires an inter-region recovery process for two reasons: first, if a node $N_x$ in $R_i$ detects a fault against $N_y$ in an upstream region $R_j$, then $R_j$ needs to be notified about the fault and both regions need to perform their local recovery; and second, if the local recovery process in a region selects a new replica for an inter-region task $\tau$, then all external regions with which $\tau$ communicates need to be informed of $\tau$'s new replica. We call this process \emph{recovery propagation (RP)}. 

\vspace{.25ex}
\bsection{The mechanism}
To notify another region $R_j$ about a fault detected in $R_i$, each measurer $N_x \in R_i$ should prepare an RP message $m_{\mathsf{rp}}$, containing the fault type, faulty node/link, affected task and job ID, and any new assignment of inter-region task replicas to be done by $R_i$'s local recovery.

Let $t_{\mathsf{rls}}^{j}$ be the release time of the affected job $j$, which will be included in the fault declaration when a fault is detected.
Let $[D_{\mathsf{det}}-\Delta_{\mathsf{det}}, D_{\mathsf{det}}]$ be the range of time taken to detect a fault relative to $t_{\mathsf{rls}}^{j}$, and 
$n$ be the smallest integer
that satisfies $t_{n}^s \geq t_{\mathsf{rls}}^{j}+D_{\mathsf{det}}$.\footnote{The value   $D_{\mathsf{det}}$ for intra-region faults is provided offline by \rebound. The maximum time  $D_{\mathsf{det}}$ to detect an inter-region fault can also be computed offline, as \system's mechanisms of inter-region fault detection have a timeout built-in (PoC timeout of commission faults, and message timeout for omission faults). As both  $D_{\mathsf{det}}$ and $t_{\mathsf{rls}}^{j}$ are fixed values determined offline, all correct nodes will agree on the same $n$.}
Then, in the measurement round $n$,
 $N_x$ should include $m_{\mathsf{rp}}$ in its signature exchanged with other measurers---i.e., send $\sigma_x(n,m_{\mathsf{rp}},\text{PoCs})$ instead of simply $\sigma_x(n,\text{PoCs})$)---and attach $m_{\mathsf{rp}}$ to the heartbeat for round $n$ to region $R_j$. Each measurer $N_y$ in $R_j$, upon receiving a heartbeat from $R_i$, will verify the signatures in the heartbeat. If they are valid signatures on $(n,m_{\mathsf{rp}},\text{PoCs})$, $N_y$ will broadcast the heartbeat to all nodes in $R_j$. These nodes will in turn verify the heartbeat and perform recovery based on the RP message $m_{\mathsf{rp}}$ if the signatures are invalid, or declare a commission fault against the sender $N_y$ otherwise.




\vspace{.25ex}
\bsection{The bounded-time recovery guarantee}
After a fault occurs, we guarantee that the entire system will resume normal operation or switch to safe mode 
within \emph{bounded time}. A proof of this BTR guarantee can be found in Appendix~\ref{sec:appendix-btr-property}.

\begin{theorem}[BTR Guarantee] \label{thm:rt-prop}
      If a fault is detected at $\tdetect$, then all nodes in the system that need to perform recovery will start the recovery or switch to a safe mode by
     $\tdetect+D_{\mathsf{RP}}$ where 
     $D_{\mathsf{RP}} = 
     2 \cdot (\Delta_{\mathsf{det}}+\tinterval+ 2\tintra + E_{\mathsf{hb}} + \thbtimeout)
     $
     if no further fault occurs in $[\tdetect, \tdetect+D_{\mathsf{RP}}]$.   
\end{theorem}

\noindent Let $\recoveryboundintra$ denote the maximum time for completing recovery actions within a region.
Theorem~\ref{thm:rt-prop} implies that if a fault is detected at $\tdetect$, the system will \emph{complete} recovery or be in a safe mode by $\tdetect + D_{\mathsf{RP}} + \recoveryboundintra$, if no further fault happens in $(\tdetect, \tdetect + D_{\mathsf{RP}} + \recoveryboundintra ]$.
If another fault does occur during this period, say at $\tdetect'$, the system is guaranteed to recover from both faults by $\tdetect'  + D_{\mathsf{RP}} + \recoveryboundintra$. Recall that there are at most $F$ faulty nodes in the entire system, so the system can still recover in bounded time, even under consecutive faults.

\section{Security Analysis and Discussion}
\label{sec:discussion}

\bsection{Safety} \label{subsec:discuss-guarantees}
\system guarantees that the system will be in a safe state within a bounded time after a fault occurs.

\vspace{.25ex}
\noindent \textit{Proof.} \system incorporates existing BTR solutions for all faults that occur within and affect a single region.
For inter-region \emph{commission faults}, Theorem~\ref{thm:consensus-poc} ensures that all correct nodes will reach agreement on whether an inter-region message is valid, and whether a safe mode should be triggered, by the timeout of the heartbeat containing the PoC for that message. 
Additionally, \system requires correct nodes to forward received inter-region messages to their peer replicas (see the ``Downstream task replicas'' paragraph under ``Fault detection'' in Sec.~\ref{subsec:inter-region-comm}). As a result, either all or none of the correct downstream replicas will receive the message, and they will collectively agree on whether a safe mode should be triggered due to message delays or \emph{omission faults}. 
Finally, in cases where a fault is detected in one region but requires recovery actions in another, Theorem~\ref{thm:rt-prop} guarantees that \emph{recovery propagation} will complete within bounded time. 
Thus, \system always guarantees bounded-time recovery from faults and ensures the system reaches a safe state within bounded time. Importantly, this holds independently of our network latency estimation accuracy and Assumption~\ref{assume:jitter}. 


\bsection{Parameter selection}
\noindent\system introduces several configurable parameters. 
Timing parameters, such as $\tinterval$ and $\thbtimeout$, should be chosen such that 
$D_{\mathsf{RP}} + \recoveryboundintra \leq \recoverybound$
where 
$D_{\mathsf{RP}}$ is defined in Theorem~\ref{thm:rt-prop} and $\recoveryboundintra$ is the maximum time needed for local recovery. 
This ensures that the system can be recovered within the recovery bound $\recoverybound$.
To choose TGS parameters—such as $\alpha$, $\beta$, and ($\pnorm$, $\Dinter$)—in practical deployments, operators should first determine $\pnorm$ and $\Dinter$ with profiled network conditions, and then run simulations as in Sec.~\ref{subsec:eval-reputation} to ensure the parameters offer strong robustness.

\bsection{Implications of network faults and attacks} \label{sec:dos-discussion}
\noindent Inter-region network faults and attacks are out of scope, but we discuss below their implications to \system.
The security properties of \system---consensus on  latency, consensus on correctness, and bounded-time recovery  (Theorems~\ref{thm:agreement}, \ref{thm:consensus-poc} and~\ref{thm:rt-prop})---are independent of the estimated network latency bound. Thus, \system guarantees safety even under network faults and attacks.
As for operational robustness, 
network faults and attacks might undermine the probabilistic bounded jitter assumption in \system: $\pnorm$ may decrease and $\Dinter$ may increase. This can cause a higher probability of over-estimating the network latency (Theorem~\ref{thm:meas-accuracy}), and further lead to more intentional delays that escape TGS suspicion. Also, more correct nodes may be flagged due to increased network delays. However, our evaluation shows that TGS is robust to network attacks to a certain degree---it remains effective with appropriate parameter configurations (see Appendix~\ref{app-subsec:tgs-eval}).



\section{Evaluation} \label{sec:eval}

\noindent We evaluated \system using a series of micro-benchmarks and case studies. Our experiments ran on machines with AMD EPYC 7452 CPUs at 2.4 GHz and 128 GB memory. Our goal is to evaluate 1) TGS's effectiveness in preventing faulty nodes from triggering a safe mode, and 2) \system's resource efficiency and effectiveness.

\begin{figure}[t]
\centering
  \begin{subfigure}{0.462\linewidth}
  \centering
    \includegraphics[width=\linewidth]{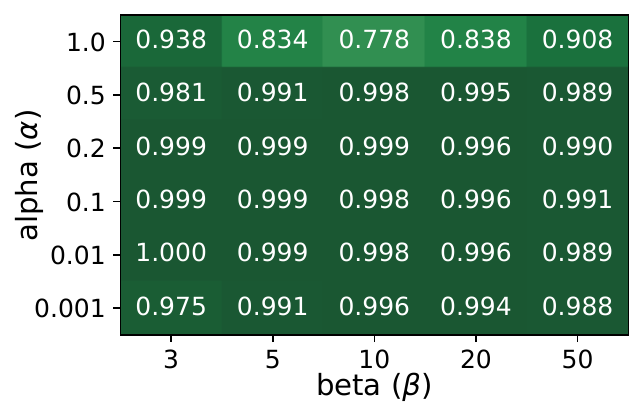}
    \caption{Adaptive attack.} 
  \end{subfigure}
  \begin{subfigure}{0.524\linewidth}
    \centering
    \includegraphics[width=\linewidth]{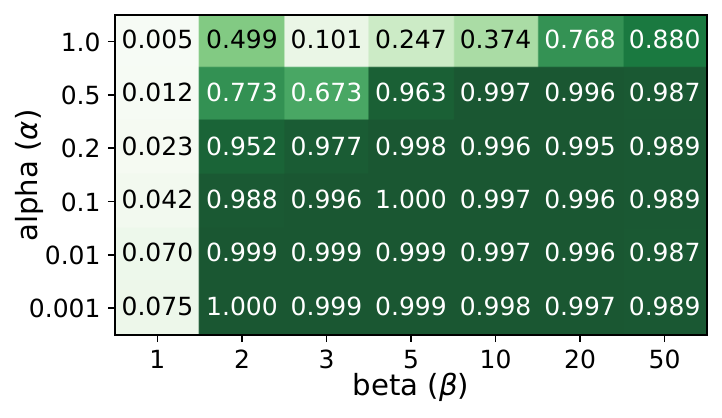}
    \caption{Aggressive attack. }
  \end{subfigure}
  \caption{Probability of staying in normal operation with TGS when $f_i=1$.  (Probability without TGS is 0.107)} \label{fig:tgs-effective}
\end{figure}

\subsection{TGS effectiveness}\label{subsec:eval-reputation}

\bsection{Setup} To evaluate TGS's ability to prevent faulty nodes from triggering safe mode transitions, we considered a system with two regions, $R_1$ and $R_2$, that execute inter-region tasks $\tau$ (upstream) and $\tau'$ (downstream), respectively.  We considered different values of $f_i$. For each value of $f_i$, we set the number of nodes in $R_i$ to be $n_i = 2f_i + 1$ and the number of replicas per task be $f_i+1$. The task replicas are initially assigned to nodes within the region at random. We set the tasks' periods and  timeouts (to trigger a safe mode) equal to 1 s. We set $\pnorm=0.999$ and varied the values of $\alpha$ and $\beta$. Recall that high $\alpha$ values impose more penalty on suspicious behaviors, whereas high $\beta$ values allow a node to exhibit more consecutive suspicious behaviors before being flagged. For each configuration of $(\alpha, \beta)$, we ran 10,000 simulations, each containing 2.59 million task invocations, which correspond to a 30-day period. We computed the probability of the system {\em not} switching to the safe mode during this period. 

We considered two attack scenarios: an \emph{aggressive} attack, in which the compromised node drops all inter-region messages whenever it is assigned to run an inter-region task; and an \emph{adaptive} attack (inspired by the attack in~\cite{DBLP:conf/iccps/NautaSM25}), in which the compromised node either (i) drops the inter-region messages, if its resulting timeliness score remains positive, or (ii) does not drop any messages, otherwise. (By the definition of $\beta$, a node will be flagged if it drops $\beta$ or more consecutive messages; thus, the adaptive attack is not applicable under $\beta \leq f_i+1$ where the node would never drop any messages.)

\vspace{.25ex}
\bsection{Results} Fig.~\ref{fig:tgs-effective} shows the probability of the system always staying in normal operation when using TGS for $f_i = 1$. The higher the probability, the more robust the system is. The results show 
that, with appropriate selection of $(\alpha, \beta)$ parameters, TGS is highly effective against both attacks.

Interestingly, an overly strict bound, with very high $\alpha$ (e.g., $\alpha=1$) or very low $\beta$ (e.g., $\beta=1$), can lead to poor TGS performance (low probability). With further investigation, we found that such a bound caused frequent false positive flags against correct nodes, resulting in frequent task reassignments and a higher frequency of the compromised nodes being selected to perform inter-region tasks.  Additionally, we noticed scenarios in which the attacker compromised a node that was initially not handling any inter-region task, but the node was later selected to run an inter-region task because TGS flagged a correct node. 
These behaviors also illustrate that a na\"{i}ve approach that flags a node whenever it is suspected of dropping/delaying a single message is not desirable.  
We further observe that an overly loose bound (e.g., $\alpha=0.01$ and $\beta=50$) also leads to poor performance. We refer the readers to Appendix~\ref{app-subsec:tgs-eval} for  detailed results and  other settings (e.g., $f_i>1$ and under \emph{DoS attacks}).

\begin{table}[t]
\centering
\scalebox{0.75}{
    \begin{tabular}{|c|c|cccc|}
    \hline
    \multirow{2}{*}{\textbf{System}} &
      \textbf{\system} &
      \multicolumn{4}{c|}{\textbf{\pistis}} \\ \cline{2-6} 
     &
      $\mathbb{T}=d$ &
      \multicolumn{1}{c|}{$\mathbb{T} = 8d$} &
      \multicolumn{1}{c|}{$\mathbb{T} =12d$} &
      \multicolumn{1}{c|}{$\mathbb{T} = 14d$} &
      $\mathbb{T}= 16d$ \\ \hline
    \textbf{Probability} &
      0.999 &
      \multicolumn{1}{c|}{0.000} &
      \multicolumn{1}{c|}{0.534} &
      \multicolumn{1}{c|}{0.978} &
      0.999 \\ \hline
    \end{tabular}
}
\caption{Comparing operational robustness against \pistis.}
\label{tbl:robust-vs-pistis}

\end{table}

\bsection{Comparison against \pistis}
We compared \system's performance against \pistis's~\cite{DBLP:journals/tpds/KozhayaDRV21},
a recent real-time Byzantine fault tolerance protocol for unreliable networks. 
For \system, we chose $(\alpha=5,\beta=0.01)$, which offers good robustness under both types of attacks.
For \pistis, we chose $f=2f_i=2$ to ensure the same total number of faulty nodes as in \system. Additionally, we let the heartbeats in each round be sent to $X=f+1$ nodes, as recommended in~\cite{DBLP:journals/tpds/KozhayaDRV21}, and varied the ratio of timeout to heartbeat period (denoted by $\mathbb{T}/d$ in~\cite{DBLP:journals/tpds/KozhayaDRV21}). Note that the configuration of \system implies $\mathbb{T}=d$. 
Table~\ref{tbl:robust-vs-pistis} shows the probability of staying in normal mode after 30 days for both systems.
With  $\mathbb{T}=8d$ (recommended by the authors), the system eventually switched to safe mode in all of 10,000 simulations. To achieve similar robustness to \system's, \pistis requires $\mathbb{T}=16d$, which leads to a much higher CPU requirement (Sec.~\ref{subsec:eval-resource}). This $\mathbb{T}$ value also makes the worst-case time for \pistis to respond to a fault $48$ times the heartbeat period (i.e., $3\mathbb{T}$, as shown in~\cite{DBLP:journals/tpds/KozhayaDRV21}) and thus significantly limits \pistis's practicability. In contrast, the recovery bound of \system, as provided in Theorem~\ref{thm:rt-prop}, is just over twice the heartbeat period.

\begin{table}[t]
            \centering
            \begin{adjustbox}{scale=1}
            \footnotesize
            \begin{tabular}{|c|c|c|c|c|}
            \hline
            \textbf{Number of Regions} & \textbf{5} & \textbf{10} & \textbf{20} & \textbf{50} \\ \hline
            \textbf{Intra-Region (kB/s)}   & 6.25       & 12.49       & 24.90       & 61.95       \\ \hline
            \textbf{Inter-Region (kB/s)}   & 7.28       & 14.54       & 29.01       & 72.00       \\ \hline
            \end{tabular}
            \end{adjustbox}
            \caption{Per-node bandwidth usage by \system in the railway control system simulation.}
            \label{tbl:bandwidth}
\end{table}

\subsection{\system resource efficiency}\label{subsec:eval-resource}
\bsection{Bandwidth requirement for \system}
To show the network bandwidth cost of \system, we ran a railway control simulation (ETCS~\cite{etcs3-spec}) with \system implemented 
in the \texttt{ns-3} simulator. The heartbeat interval $\tinterval$ was set to $1$~s. The results in Table~\ref{tbl:bandwidth} show that the bandwidth per node is on the kbps level when it scales to 50 regions, and it grows linearly with the number of regions.

\bsection{Computation efficiency for network measurement}
We compared the efficiency of \system's network latency measurement module with \msptp~\cite{DBLP:conf/wisec/ShiXDSLZ0L23}, which offers Byzantine-resilient latency measurement. Note that, unlike \system, \msptp\ {\em  does not guarantee agreement}.

For both \msptp and \system, we set up two regions and had nodes measuring the latency between them.
We varied the maximum number of faulty nodes ($f_i$) in both regions, and compared the total CPU time among all nodes to complete a single round of latency measurement. This includes the time to sign/verify and send/receive messages, along with the time to process the timestamps and compute an estimate.  We ran each protocol 1000 times under each setting, and listed the averages in Table~\ref{tbl:comparison-measurement}. Results show that 
\msptp consumes significantly more (up to 4.88$\times$) CPU resources than \system. This is expected as \msptp requires nearly 3$\times$ measurers ($3f_i+1$ instead of $f_i+1$ per region). 

\begin{table}[t]
    \centering
    \begin{adjustbox}{scale=1}
    \footnotesize
    \begin{tabular}{|c|c|c|c|c|c|}
    \hline
    \textbf{$f_i$ for both regions}   & \textbf{1} & \textbf{2} & \textbf{3} & \textbf{4} \\ \hline
    \textbf{\msptp (ms)}  & 11.30      & 30.93      & 61.28      & 102.05 \\ \hline
    \textbf{\system (ms)} & 2.36       & 6.34       & 12.93      & 22.67 \\ \hline
    \end{tabular}
    \end{adjustbox}
    \caption{Total CPU time per network measurement round.}
    \label{tbl:comparison-measurement}
\end{table}

\begin{figure}[t]
    \centering
    \includegraphics[width=.82\linewidth]{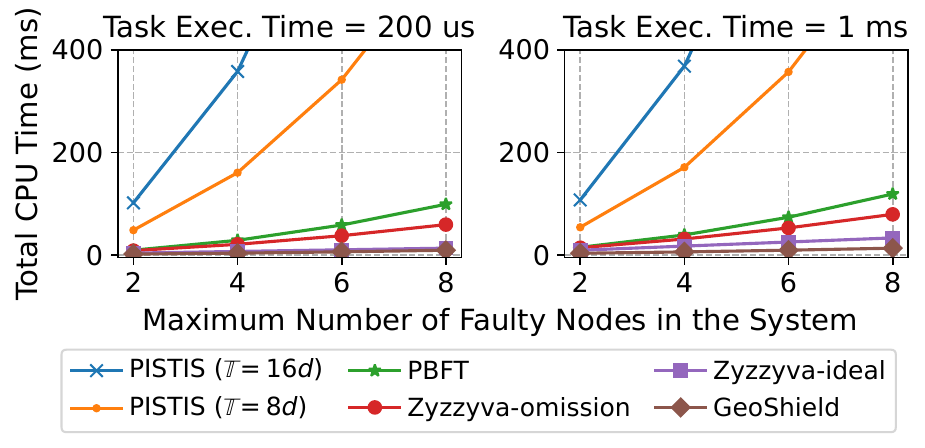}
    \caption{Total CPU time to complete a job under \system vs. existing BFT protocols.}
    \label{fig:bft-cpu-comparison}
\end{figure}

\bsection{Comparison against related solutions}
We compared \system against three BFT protocols: \pbft~\cite{castro1999practical}, \Zyzzyva~\cite{kotla2007zyzzyva}, and \pistis~\cite{DBLP:journals/tpds/KozhayaDRV21}. 
\pbft and \Zyzzyva represent the classical pessimistic and optimistic BFT solutions, respectively, while  \pistis aims to provide worst-case timing guarantees in BFT. 
For \Zyzzyva, we consider two cases: When no fault or message delay occurs (\ZyzzyvaIdeal), the job is completed when a node receives $3f+1$ identical outputs.
However, if any message does not arrive, \Zyzzyva goes to a slow path (\ZyzzyvaOmission). As \Zyzzyva has no mechanism to detect omission faults, a malicious node can easily cause \Zyzzyva to degrade to \ZyzzyvaOmission without being detected. Therefore, \ZyzzyvaIdeal is difficult to achieve in the presence of malicious nodes. In contrast, TGS in \system can detect faults and recover the system. 

We considered different values for the average execution time of a task and total number of faulty nodes ($F$) in the system. For \system, we assumed two regions with $f_1 = f_2 = F/2$ ($F$ is even). 
For \pistis, we considered two settings: $\mathbb{T}=8d$, as recommended by the authors~\cite{DBLP:journals/tpds/KozhayaDRV21}, and $\mathbb{T}=16d$, where good operational robustness can be achieved. 

For each configuration, we computed the total CPU time among all the nodes to complete a job (a task invocation), which includes the time to execute jobs, sign/verify messages, and send/receive packets. We ran each protocol 1000 times under each configuration, and reported the averages in Fig.~\ref{fig:bft-cpu-comparison}. %

{\em Results.} The results show that \system outperforms both \pbft and \Zyzzyva, with its improvements increasing as the computation load of jobs increases. This is expected, as \system needs much fewer replicas of jobs compared to BFT protocols. 
On the fast path, \ZyzzyvaIdeal uses 1.43$\times$ to 2.07$\times$ as much CPU resource as \system under a light application workload, and 2.51$\times$ to 2.89$\times$  as much under a heavier workload.
On the slower path, \ZyzzyvaOmission's CPU usage is 4.20$\times$ to 6.38$\times$ that of \system. 
\pbft and \pistis are even less efficient due to their pessimistic designs. For instance, \pistis relies on very frequent heartbeats (e.g., 8 or 16 heartbeats per job) to maintain timeliness and operational robustness, leading to substantial overhead for generating, sending, and verifying signatures. Consequently,  \pistis requires significant more CPU resources (16.0$\times$ to 151.8$\times$) compared to \system. 

\begin{figure*}[t!]
    \begin{minipage}{0.58\linewidth}
        \centering
        \includegraphics[width=1\linewidth]{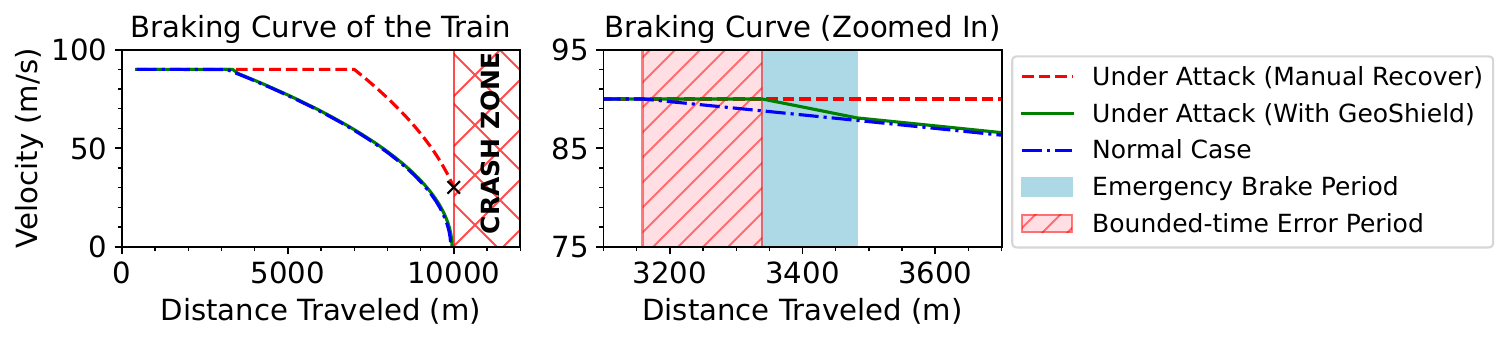}
        \caption{Braking curve under incorrect MA attack.}
        \label{fig:brake-curve}
    \end{minipage}
    \begin{minipage}{0.4\linewidth}
    \centering
        \includegraphics[width=.75\linewidth]{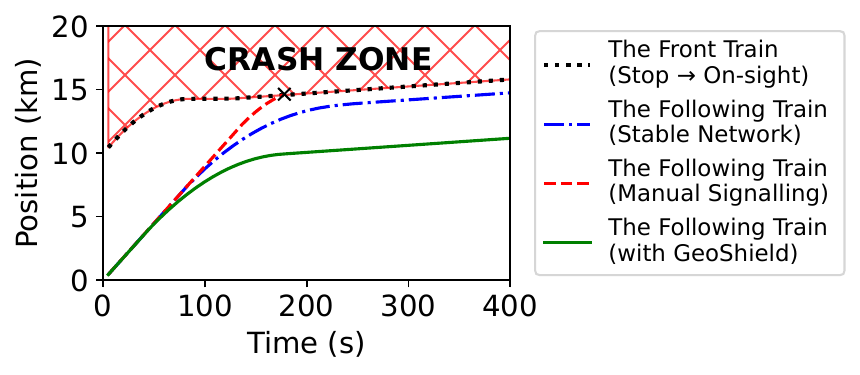} 
        \caption{Trains' locations with lossy networks.}
        \label{fig:train-locations}
    \end{minipage}
\end{figure*}

\subsection{Case study: Railway control system}
\noindent We now present an attack against a railway
control system running ETCS-3~\cite{etcs3-spec} to evaluate the effectiveness of \system. ETCS and its variants are widely deployed in Europe, China, India, Australia, etc.
We used \texttt{ns-3} to simulate the train(s) and the control center with \system deployed, as well as the communications between them.
The communications go through an LTE network with eNBs (the base stations) alongside
the track for every 10~km. The trains perform automatic
handover to connect to the nearest eNB when moving. 
Under ETCS, the control center sends movement authorizations (MA) to trains, with
information about how far the train can move. The train stores the MA and
runs an onboard control algorithm that ensures it never goes beyond the MA,
unless the MA is renewed.
ETCS also specifies an emergency brake and a service
brake with a smaller deceleration to maximize the comfort of passengers. We
set the deceleration of the emergency brake and service brake to be 1.2~$\text{m}/\text{s}^2$ and 0.6~$\text{m}/\text{s}^2$, respectively.
The heartbeat period $\tinterval$ was set to $1$~s and the timeout $\thbtimeout$ set to $200$~ms, which ensure the recovery bound (by Theorem~\ref{thm:rt-prop}) is within the ETCS specification~\cite{etcs3-spec,finnish-etcs-values,itea-etcs-values}. 

\noindent\textbf{Incorrect movement authorization attack.} 
Consider a train at location $x_0=0$ of a track, moving at 90~m/s, and an obstacle at $x_{\max}=10$~km. Consider a node in the control center is compromised, and it replaces the MA by $x_{\max}' = 20$~km. We assume the driver can see the obstacle no more than 3 km away, and she can start the emergency brake thereafter.

Fig.~\ref{fig:brake-curve} shows the braking curve of the train in three
different scenarios. In the normal case, the train
starts a service brake when it travels to $x=3.16$~km ($t=35.10$~s) and it 
 reaches a full stop at $x=9.91$~km. 
In the attacked scenario, the train follows the incorrect MA and does not start braking automatically. 
The driver sees the obstacle at $x=7.00$~km and starts the emergency braking immediately, but the train still crashes.
With \system, when the train receives the MA at $t=35.10$~s,
it expects a PoC attached to the heartbeat message scheduled at $t=36.00$~s. 
Since the MA is incorrect, no valid PoC with enough endorsements on that MA will be generated.
The nodes on the train declare a fault against the sender when they do not see a valid PoC in the heartbeat, and propagate it to the control center. The nodes in the control center send a corrected MA with $x_{\max}=10$~km, which arrives at $t=36.08$~s ($x=3.247$~km).
The train cannot stop in time with the service brake, so it experiences
an emergency brake over a short period (the hatched area in Fig.~\ref{fig:brake-curve}-right) and switches to a service brake afterward. 
This example shows that \system can ensure safety with bounded-time recovery. 

\bsection{Train collision incident in Wenzhou, China, 2011}
Next, we show that \system can ensure safety in a multi-train scenario with bounded-time recovery propagation. We use the incident
in Wenzhou, China, 2011 (discussed in the introduction) as a use case. 
Fig.~\ref{fig:train-locations} shows the simulated positions of the two trains.
Two trains have an initial distance of 10~km, both moving at 90~m/s. At
$t=10.0$~s, lightning strikes the front train, so it loses communication
with the control center and starts a forced braking. At $t=120.0$~s, the driver
takes over and moves the train at $20$~km/h in the on-sight (manual) mode.
If there had been good network connectivity (the blue line), the second train could have switched to on-sight mode and slowed down in time.
In that incident (the red line), the
driver received a manual communication from the control center at $t=120.0$~s, but it
was too late. 
With \system (the green line), the control center expects a heartbeat
message scheduled at $t=10.0$~s (which cannot arrive due to the fault), and at $t=10.2$~s, a safe mode is
triggered, in which case the control center will update the MA of the following train to be the front train's prior location to ensure safety.

\subsection{Case study: Smart grid}
\begin{figure}[t!]
    \centering
        \includegraphics[width=.9\linewidth]{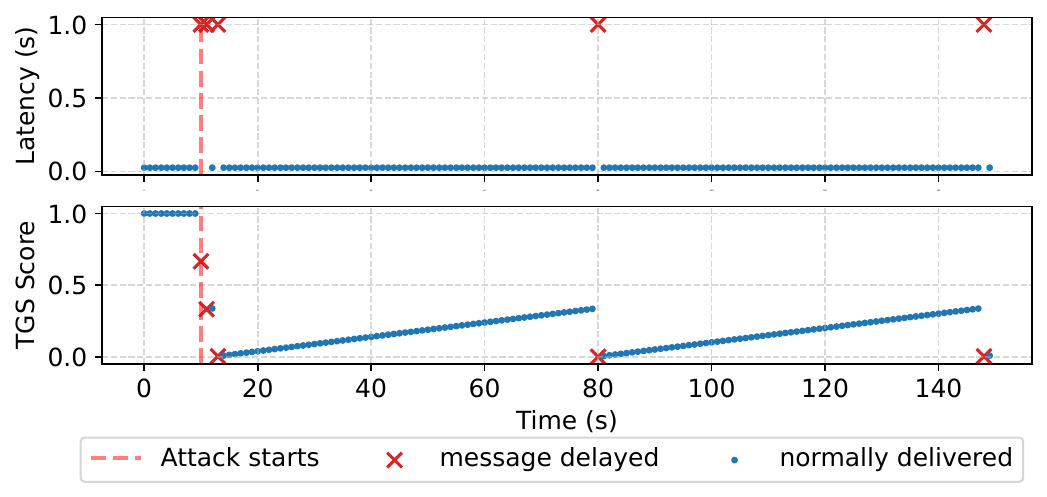}   
    \caption{The query response latency (top) and the scores (bottom) of the nodes in a smart grid.}
    \label{fig:faulty-grid}
\end{figure}

\noindent We next demonstrate how \system can enhance operational robustness in a smart grid system. 
We used the workload and data from~\cite{DBLP:conf/debs/MutschlerLWEP14} for real-time monitoring of power usage, where
An anomaly detection algorithm runs periodically in each substation. If an unusual voltage is detected, different substations need to coordinate to perform fault isolation, such as shutting down the power switches between them to avoid equipment damage.  
We set up two virtual substations in UT and WA. From our experimental validation (Sec.~\ref{sec:jitter-property}, Fig.~\ref{fig:latency}), we selected $\Dinter = 2.406$~ms and $\pnorm=0.999$. We set $\beta=3$ and $\alpha=0.2$. The attacker compromises a node at $t=10$s, and adopts the adaptive strategy for delaying messages by 1s.  
Fig.~\ref{fig:faulty-grid} shows the latency of each query response and the change in the node's score. The compromised node delays three messages in a short period, but then its score becomes low and stops doing so to avoid being flagged. Later, the node can delay only one of every 66 messages. 
This shows that TGS can effectively enhance operational robustness.

\section{Related Work}


\bsection{CPS security}
In CPS, timeliness is a critical security property alongside execution integrity and has been considered in existing security solutions~\cite{li2024data,gandhi2025roborebound}, regarding scheduler, I/O~\cite{DBLP:conf/sp/WangLL0022}, and memory corruption~\cite{li2025software} for instance.
\system aims to address the challenge of unreliable networks.

\bsection{Byzantine fault tolerance}
There is rich body of literature on BFT protocols~\cite{castro1999practical,
kapitza2012cheapbft,
DBLP:conf/rtas/BohmDW24,DBLP:conf/podc/YinMRGA19}, with varying focuses and trust assumptions, some of which targeted at CPS~\cite{bommareddy2022real,cheng2021fault}. As BFT protocols mask all faults, they require more replicas and communication rounds than BTR. Among these, \pistis~\cite{DBLP:journals/tpds/KozhayaDRV21} has the closest goal to ours---providing worst-time timing guarantees under unreliable networks. 
However, \system offers significantly better efficiency and operational robustness.




\bsection{Bounded-time recovery}
Different recovery approaches for CPS have been proposed for crash faults~\cite{Gandhi2020BoundedtimeRF,shu2002achieving} and certain types of attacks and systems (e.g., misconfiguration~\cite{DBLP:conf/raid/Zhang00M22} and stochastic CPS~\cite{DBLP:conf/rtas/ZhangBCC0AK24}, respectively). The idea of bounded-time recovery (BTR) was proposed in~\cite{chen2015fault}, and a solution for Byzantine faults under fully synchronous, reliable networks was developed by \rebound~\cite{gandhi2021rebound}. More recently, \roborebound~\cite{gandhi2025roborebound} extended  \rebound~\cite{gandhi2021rebound} to a multi-robot setting with unreliable networks by leveraging trusted hardware. \system is the first BTR solution against Byzantine faults for unreliable networks without relying on trusted hardware. 

\bsection{Latency measurement and delay localization} 
\msptp~\cite{DBLP:conf/wisec/ShiXDSLZ0L23} and \texttt{ARCHER}~\cite{eischer2018latency} also enable network latency estimation in the presence of Byzantine nodes. However, \msptp does not guarantee agreement on the latency, while \texttt{ARCHER} cannot prevent faulty nodes from causing over-estimation. Therefore, they are unsuitable for TGS in \system. 
Other techniques that aim to detect Byzantine behaviors (e.g., grey-hole attacks) in networked systems use a trusted party to check the traffic patterns~\cite{liu2013fade,khalil2010unmask,zhang2012secure} or assume a fixed and known threshold on the delay or loss~\cite{curtmola2008bsmr,DBLP:conf/conext/ArgyrakiMS10}. 
In contrast, we target settings where the network is dynamic and trusted nodes are not available.


\section{Conclusion}
\noindent We presented \system, a BTR solution for geo-distributed CPS with unreliable inter-region networks.  
With its Byzantine-resilient measurement protocol,
\system effectively detects inter-region faults. Its recovery protocol ensures that the system recovers within a bounded time while maintaining operational robustness.
Evaluation results show that \system effectively protects the system against Byzantine faults, is suitable for deployment in real-world CPS scenarios to enhance security, and significantly outperforms existing BFT solutions in resource efficiency and robustness.

\cleardoublepage
\appendix

\bibliographystyle{plain} 
\bibliography{references}

\newpage
\appendices
\section*{Appendices}

\section{Additional measurements and evaluations}
\subsection{Additional results for inter-region network jitter measurement study}

\label{app:jitter-meas}
\begin{figure}[h]
\includegraphics[width=.48\linewidth]{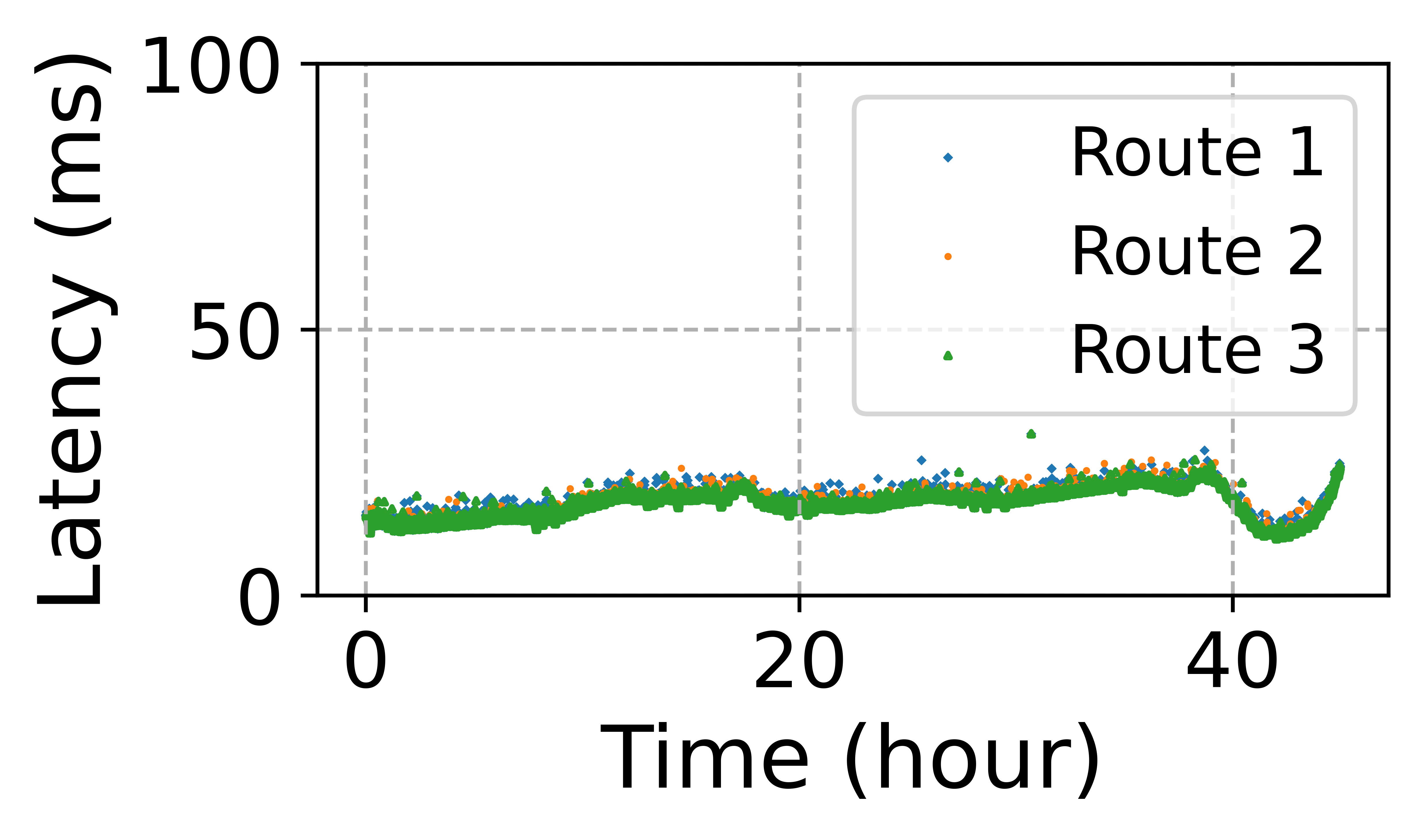}
\includegraphics[width=.48\linewidth]{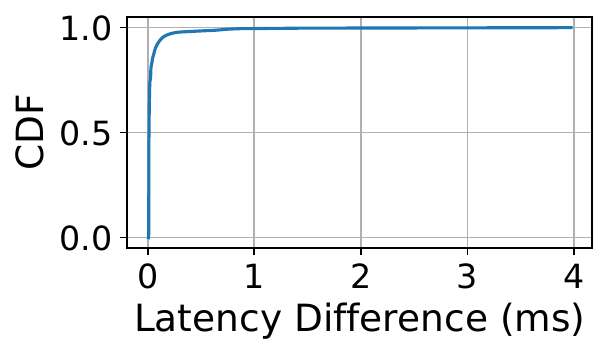}
\caption{Latency of three example routes from UT to WA (left) and the CDF of their latency difference (right).}\label{fig:latency}
\end{figure}
\noindent To further demonstrate the probabilistic bounded jitter property in Sec.~\ref{sec:jitter-property}, 
we show  an example of the latency from UT to WA in about 48 hours and the CDF of latency difference between the routes in Fig.~\ref{fig:latency}. We notice that even though the latency between the two regions may change over time, the latency difference of successive packets is small with a high probability. For instance, the 99th percentile difference is only 0.694~ms from UT to WA. 
\begin{figure}[t]
\centering
  \begin{subfigure}{0.78\linewidth}
    \includegraphics[width=\linewidth]{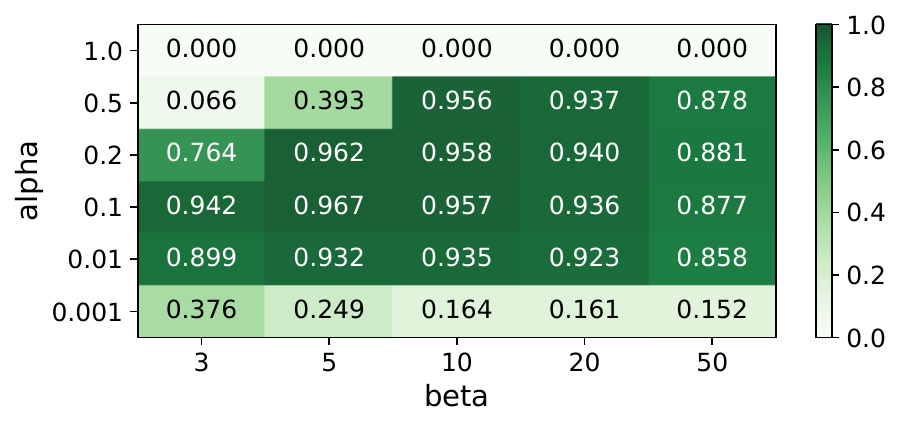}
    \caption{Adaptive attack (without TGS: 0.105)}
  \end{subfigure}
  \begin{subfigure}{0.78\linewidth}
    \includegraphics[width=\linewidth]{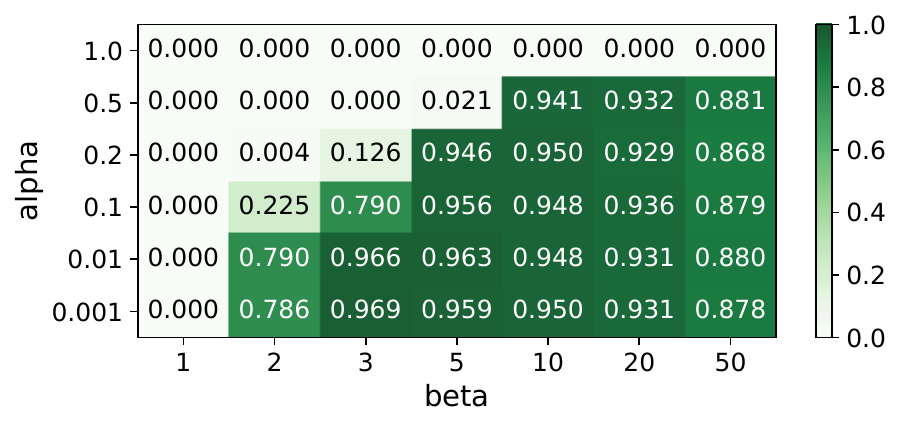}
    \caption{Aggressive attack (without TGS: 0.105)}
  \end{subfigure}
  \caption{TGS effectiveness with $\pnorm=0.99$ and $f_i=1$} \label{fig:tgs-effective-99norm}
\end{figure}
\begin{figure}[h]
\centering
  \begin{subfigure}{0.78\linewidth}
    \includegraphics[width=\linewidth]{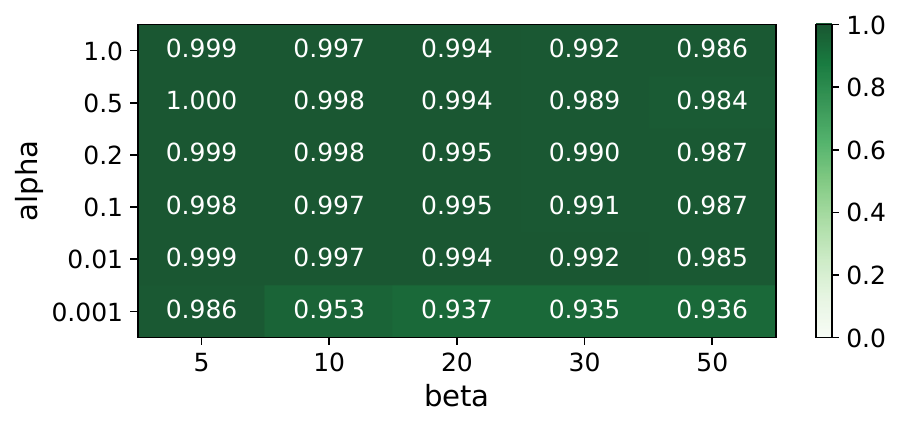}
    \caption{Adaptive attack (without TGS: 0.486)}
  \end{subfigure}
  \begin{subfigure}{0.78\linewidth}
    \includegraphics[width=\linewidth]{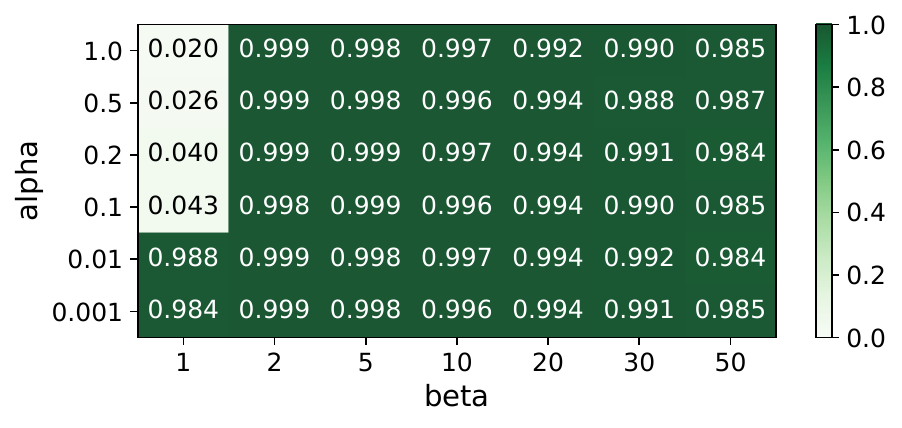}
    \caption{Aggressive attack (without TGS: 0.486)}
  \end{subfigure}
  \caption{TGS effectiveness with $\pnorm=0.99$ and $f_i=2$} \label{fig:tgs-effective-f2}
\end{figure}
\begin{figure}[h]
\centering
  \begin{subfigure}{0.78\linewidth}
    \includegraphics[width=\linewidth]{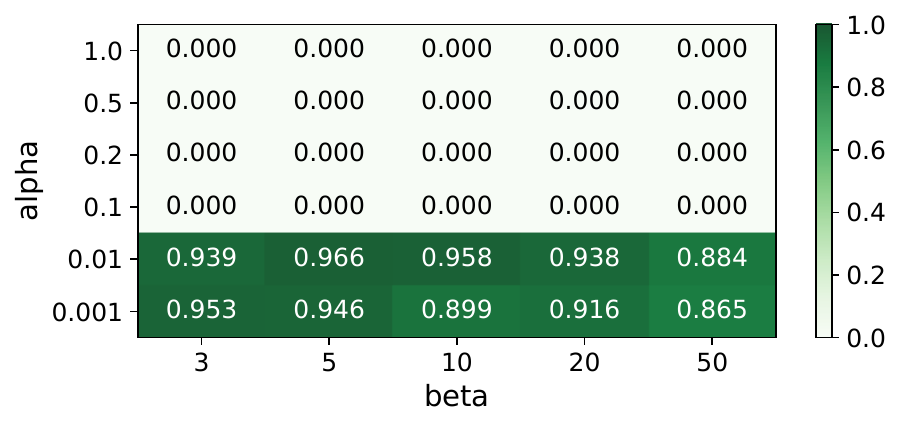}
    \caption{Adaptive attack (without TGS: 0.105)}
  \end{subfigure}
  \begin{subfigure}{0.78\linewidth}
    \includegraphics[width=\linewidth]{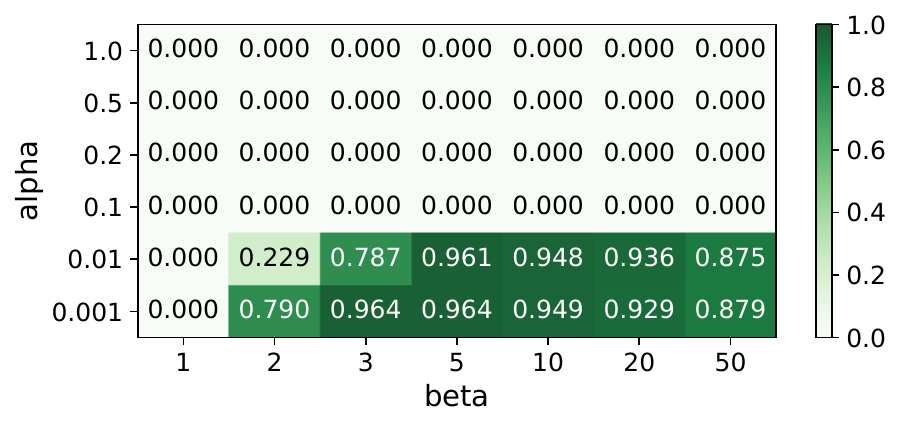}
    \caption{Aggressive attack (without TGS: 0.105)}
  \end{subfigure}
  \caption{TGS effectiveness under network DoS ($f_i=1$)} \label{fig:tgs-effective-dos}
\end{figure}

\subsection{Additional evaluation results on TGS} \label{app-subsec:tgs-eval}
\noindent This section presents additional evaluation results on the effectiveness of TGS in maintaining operational robustness. 

\bsection{Performance under different network conditions}
Fig.~\ref{fig:tgs-effective-99norm} shows the effectiveness of TGS under $\pnorm=0.99$ and $f_i = 1$, using the same $\Dinter$ and experiment setup as in Sec.~\ref{subsec:eval-reputation}. Compared to the case $\pnorm= 0.999$ that was discussed in Sec.~\ref{subsec:eval-reputation}, a network with $\pnorm=0.99$ is more unreliable; thus, maintaining the system's operational robustness is also more challenging. However, our evaluation results show that, by carefully choosing the parameters (e.g, $\alpha=0.1, \beta=5)$, TGS is able to maintain the system in normal mode for a month-long period with a probability close to one ($\geq 0.956$) under both attack scenarios. The results further demonstrate that extreme values---too large $\alpha$ or too small $\beta$---reduce the effectiveness of TGS, which is consistent with our observation under $\pnorm= 0.999$.
It is worth noting that the \emph{long-term} bound guaranteed by $\alpha$ is not effective against the \emph{aggressive} attack, as this attack drops all messages, and thus setting a very small $\alpha$ does not undermine the effectiveness of TGS. 
However, this is not the case for the adaptive attack. Since the adversary can perform any attack strategy, operators should select $\alpha$ and $\beta$ carefully to achieve the best performance across different attack scenarios.

\bsection{Performance under different numbers of faulty nodes}
Fig.~\ref{fig:tgs-effective-f2} shows the effectiveness of TGS, with $\pnorm=0.99$ and $f_i=2$. As $f_i$ increases, the number of replicas per task also increases, which in turn increases the probability of having at least one message per task output delivered in time. Consequentially, the probability of the system remaining in the normal mode is also higher with or without TGS (the probability is 0.486 without TGS). 

The results show that TGS continues to be highly effective in this setting. By choosing $\alpha=0.01$ and $\beta=5$, for instance, TGS can keep the system in normal mode for over a month-long period with probability of 0.998 against {\em both} attack strategies. Overall, we also observe substantially higher probability values under $f_i = 2$ compared to the case $f_i = 1$, across all settings of $\alpha$ and $\beta$.

\bsection{Performance under DoS attacks on the network}
As discussed in Sec.~\ref{sec:dos-discussion}, network attacks might undermine the probabilistic bounded jitter assumption, leading to an over-optimistic estimate of $\pnorm$ and consequentially a higher probability of our network measurement protocol over-estimating the network latency. To investigate the impact of network attacks on TGS, we considered the scenario in which the inter-region network is under a DoS attack, using the same experimental setup as in the previous experiments. We assumed the same original estimate of $\pnorm=0.999$ for the inter-region network (based on our experimental validation of the network latency jitter). However, at run time, we assumed that a DoS attack degraded the network, bringing the actual $\pnorm$ down to only $0.99$. This attack effectively enlarged the jitter rate by 10 times compared to our jitter assumption. 

Fig.~\ref{fig:tgs-effective-dos} shows the probability of TGS maintaining the system in normal mode under DoS attack for the case where $f_i=1$. The results show that, with careful parameter selection, TGS remains effective. For example, consider all 6 scenarios that were evaluated, TGS with $\alpha=0.01,\beta=5$ can maintain the system in normal operation for over a month-long period with probability of 0.932. 

Note that we did not evaluate the cases where the attack makes $\pnorm$ even smaller (e.g., 0.9 or less). For those cases, the attacker can force the system to switch to a safe mode easily even in the absence of any Byzantine nodes, and thus is out of the scope of \system.

\section{Additional details on the network latency measurement protocol}
\label{app-sec:meas}
In this appendix, we provide more details on the network measurement protocol presented in Sec.~\ref{sec:measurement}. We start with the proof of the accuracy of measurement. We then present the detailed protocol for fault detection and recovery during the measurement process. Based on the protocol, we can then prove that all faults during the process can be detected in bounded time, and that all correct nodes will agree on the same latency value. 

\subsection{Proof of measurement accuracy}
\vspace{1ex}
\noindent{\bf Lemma \ref{lem:no-early} (Early heartbeat).}
{\em     It is impossible for a node to send a valid heartbeat message more than $\tearly = \Dsyn + \Dintra + \Dhb$  time units before the scheduled time. 
}     
\begin{proof}
    Recall that $\Dsyn$, $\Dintra$ and  $\Dhb$ are the maximum clock skew, the maximum intra-region network latency jitter, and the maximum difference in the time a node takes to verify signatures and construct a heartbeat, respectively. 
     
    Consider a valid heartbeat from node $N_x$ in region $R_i$. This heartbeat must contain $f_i+1$ signatures, one of which must be from a correct node $N_y$ (since at most $f_i$ nodes can be faulty in $R_i$). Since $N_y$ is correct, it sends the signature at $t_n^s$, 
    and the signature reaches $N_x$ no earlier than $t_n^s + (\tintra - \Dintra)$ according to $N_y$'s clock. As it takes at least $\thb-\Dhb$ time units to construct the heartbeat, the earliest time $N_x$ sends the heartbeat message according to $N_y$'s clock is
    $t_n^s + (\tintra - \Dintra) + (\thb-\Dhb) = (\tsendn - \tintra - \thb) + (\tintra - \Dintra) + (\thb-\Dhb) = 
    \tsendn - \Dintra - \Dhb$.
    Since the nodes' clocks differ by at most $\Dsyn$, node $N_x$ sends the heartbeat no earlier than 
    $\tsendn - \Dintra - \Dhb - \Dsyn = \tsendn -\tearly$ according to any node's clock. Hence, the theorem.
\end{proof}

\vspace{.25ex}
\noindent {\bf Theorem~\ref{thm:meas-accuracy} (Accuracy). }
{\em Suppose the actual probabilistic maximum inter-region latency in round $n$ is $D_{\mathsf{real,n}}$, then

    $\tdecide \geq D_{\mathsf{real,n}} -  (\tearly + \Dprop + \Dintra + \Dsyn)$; and 

    $\tdecide \leq D_{\mathsf{real,n}}+ (\Dinter+\Dsyn) $, with probability $\ge \pnorm$.}

\begin{proof}
    First, consider the earliest time when a proposal is accepted. From Lemma~\ref{lem:no-early}, the earliest time any upstream measurer can send a valid heartbeat is $\tsendn - \tearly$. Since the actual probabilistic maximum inter-region network delay is $D_{\mathsf{real,n}}$, the heartbeat takes at least $D_{\mathsf{real,n}}-\Dinter$ time units to reach a downstream measurer $N_x$, where $\Dinter$ is the probabilistic jitter bound (see Assumption~\ref{assume:jitter}). The measurer $N_x$ takes at least $\tprop-\Dprop$ to process the heartbeat and makes a proposal, which further takes at least $\tintra-\Dintra$  to reach another peer measurer $N_{x'}$. Thus, $N_{x'}$ receives the proposal time at $t \geq \tsendn-\tearly+(D_{\mathsf{real,n}}-\Dinter) + (\tprop - \Dprop) + (\tintra - \Dintra)$. In Phase 3a, for a proposed latency value to be considered reasonable, it must be at least $d_{\min} = t - \tsendn - \tprop - \tintra - \Dsyn
    \geq D_{\mathsf{real,n}} - (\tearly + \Dinter + \Dprop + \Dintra + \Dsyn)$. 
    Since $\taccept = \min(A_{\mathsf{min}}, P_{\mathsf{min}}) + \Dinter$, 
    we imply that the accepted value $\taccept$ and consequentially the decided value $\tdecide$ must be at least $D_{\mathsf{real,n}} - (\tearly + \Dprop + \Dintra + \Dsyn)$. 

    Next, consider the probabilistic worst case in which all heartbeats spend $D_{\mathsf{real,n}}$ in the network. Since at least one heartbeat must be from a correct measurer, that measurer must have sent it at $t_n$. Thus, the heartbeat arrives at the receiving measurers at $\tsendn + D_{\mathsf{real,n}}$, which implies that the maximum value that the measurers propose is $D_{\mathsf{real,n}} + \Dsyn$.
    Hence, the probabilistic worst-case network latency accepted is at most  $D_{\mathsf{real,n}}+\Dinter+\Dsyn$, and the decided value $\tdecide$ will be at most $D_{\mathsf{real,n}}+\Dinter+\Dsyn$ with probability of at least $\pnorm$. 
\end{proof}

\subsection{Detection and recovery during the measurement process} \label{subsec:app-detection-recovery-meas}
In this subsection, we present the detailed protocol for detection and recovery that is used during measurement. 

\bsection{Overview} At the \emph{decide} phase (Phase 4 in Sec.~\ref{subsec:meas-phases}) of the latency measurement protocol, if any of the measurers detects that another measurer has decided a different value from itself, it will declare a fault and disseminate it to all the downstream measurers and log keepers. At a high level, the measurers and log keepers will exchange their logs, detect faults based on the logs, and decide on a new latency bound.

\bsection{Stage 1: Fault declaration}
When a measurer $N_x$ receives an accept message $m_{\mathsf{acc},y}$ from $N_y$, if it detects that the $m_{\mathsf{acc},y}$ has a different accepted value than its own, it will start  preparing for a fault declaration message $m_{\mathsf{dclr},x}$. The message consists of (i) the proposal message $m_{\mathsf{prop}, x}$ corresponding to the latency that $N_x$ has accepted, and (ii) the message $m_{\mathsf{acc},y}$ from $N_y$ that $N_x$ has considered faulty. $N_x$ will then sign $m_{\mathsf{dclr},x}$ and send it to all other measurers and $f_i$ log keepers in its region. We denote the time $N_x$ sends the declaration as $t_{\mathsf{dclr}}$.

\vspace{.25ex}
\bsection{Stage 2: Declaration validation}
To prevent a faulty measurer $N_x$ from sending the declaration to only a subset of nodes, every node receiving $m_{\mathsf{dclr},x}$ will forward it to all other measurers and log keepers. Nodes keep only the first $m_{\mathsf{dclr},x}$ they receive. Thus, by time $t_{\mathsf{dclr}}+2\tintra$, all measurers and log keepers should have received the declaration. If $N_z$ receives a forwarded declaration, but receives no declaration directly from $N_x$, it declares an intra-region omission fault against $N_x$.

After forwarding the fault declaration that it receives, each receiver $N_z$ first validates whether the declaration is indeed valid. For this, it verify the signatures of $m_{\mathsf{prop},x}$ and  $m_{\mathsf{acc},y}$ in $m_{\mathsf{dclr},x}$. If the signatures are not valid, then $m_{\mathsf{dclr},x}$ is an invalid declaration, and $N_y$ will declare a commission fault against $N_x$. It then checks if the latencies in $m_{\mathsf{prop},x}$ and $m_{\mathsf{acc},y}$ are different. If they are not different, the declaration should not have been sent (as $N_x$ should only declare a fault when it receives a different accepted latency from $N_y$'s), and $N_z$ will declare a commission fault against $N_x$. We denote by $E_{\mathsf{dclr\_v}}$ the worst-case execution time of this declaration validation process.

\vspace{.25ex}
\bsection{Stage 3: Log cross-validation}
Recall that all measurers store the proposals they have sent and all valid proposals they have received in their log, and all log keepers store the received proposals along with the received time. 
Whenever $N_z$  deems a fault declaration valid (stage 2), it should exchange its log with other measures and log keepers to identify any faulty behaviors based on the log. If $N_z$ is a log keeper, before sharing its log, it should validate all the proposals in the log, in the same manner as done by the measurers in Phase 3a of Sec.~\ref{subsec:meas-phases}. $N_z$ then signs its proposal log and sends it to all other measurers and log keepers. Let  $E_{\mathsf{log\_ex}}$ be the worst-case execution time of this log exchange process. Then, by $t_{\mathsf{dclr}} + 2\tintra +  E_{\mathsf{dclr\_v}} + E_{\mathsf{log\_ex}}$, all measurers and log keepers should have sent their logs. By $t_{\mathsf{dclr}} + 3\tintra +  E_{\mathsf{dclr\_v}} + E_{\mathsf{log\_ex}}$, if $N_z$ has not received the log from some node $N_v$, $N_z$ will declare an omission fault against $N_v$.

After receiving the logs from other nodes, $N_z$ detects faults using the following procedure: First, it verifies the signature of each proposal in each node's log. If there exists a proposal in $N_v$'s log with an invalid signature, $N_z$ marks this proposal as invalid and declares a commission fault against $N_v$. 
Second, $N_z$ checks $N_y$'s log to verify if the latency value in $m_{\mathsf{acc},y}$ indeed matches the smallest proposed latency plus $\Dinter$ in $N_y$'s log. If not, $N_z$ declares a commission fault against $N_y$. 
Next, for each pair of upstream and downstream measurers $(N_{u}, N_{d})$, it checks if the valid proposals containing the heartbeat from $N_u$ to $N_d$ in all nodes' log have the same proposed latency. If not, $N_z$ declares a commission fault against $N_d$ for equivocation. 
Further, if a valid proposal containing the heartbeat from $N_u$ to $N_d$ exists in some node's log but not in $N_z$'s own log, $N_z$ declares an omission fault against $N_d$, as $N_d$ must have delayed or dropped the proposal that should have been sent to $N_z$ in time.

After finishing the checks above, $N_z$ searches for all pairs $\{(N_u,N_d)\}$ such that the proposals (with valid signatures) based on hearbeats from $N_u$ to $N_d$ appear in at least $f_i+1$ of the logs.
Among them, $N_z$ finds the pair $(N_u,N_d)$ with the smallest latency value. $N_z$ finally generates a new accept message consisting of $N_u$, $N_d$, and $f_i+1$ proposals that contain the heartbeat from $N_u$ to $N_d$. $N_z$ signs this new accept message and sends it to all other measurers and log keepers. Assuming all processes in this stage takes at most $E_{\mathsf{log\_v}}$, then by time $t_{\mathsf{dclr}} + 3\tintra +  E_{\mathsf{dclr\_v}} + E_{\mathsf{log\_ex}} + E_{\mathsf{log\_v}}$, this message should have been sent. 

\vspace{.25ex}
\bsection{Stage 4: Deciding on a new latency}
When receiving a new accept message, $N_z$ should immediately forward it to all other measurers and log keepers to detect equivocation. Therefore, by $t_{\mathsf{dclr}} + 5\tintra +  E_{\mathsf{dclr\_v}} + E_{\mathsf{log\_ex}} + E_{\mathsf{log\_v}}$, $N_z$ should have received multiple copies of new accept messages from each other measurer and log keeper. If not, $N_z$ declares an omission fault against the sender. Next, for each measurer and log keeper $N_v$, $N_z$ verifies whether all copies of its new accept messages have the same content (to detect equivocation). If not, $N_z$ marks them as invalid and declares a commission fault against $N_v$. Then, for each sender's new accept message, $N_z$ checks if it has $f_i+1$ properly signed proposals (to validate that it is a valid latency). If not, $N_z$ marks it as invalid and declares a commission fault against $N_v$. Then, among all \emph{valid} new accept messages, $N_z$ selects the one with the smallest latency $A_{\min}$, and decides on $A_{\min}+\Dinter$. $N_z$ will then sign its decided value and broadcast to all nodes in its region. Assuming this process takes at most $E_{\mathsf{decide}}$, then all nodes will have decided on a new latency by $t_{\mathsf{dclr}} + 5\tintra +  E_{\mathsf{dclr\_v}} + E_{\mathsf{log\_ex}} + E_{\mathsf{log\_v}} + E_{\mathsf{decide}}$.

\subsection{Properties}

Based on the measurement protocol (Sec.~\ref{subsec:meas-phases}) and the above   detection and recovery protocol during measurement, we can establish the  consensus on the estimated latency bound. 

\vspace{2ex}
\noindent {\bf Theorem~\ref{thm:agreement} (Consensus on latency). }
{\em The following properties hold: 
(1) At $t_{n}^{\mathsf{dec}}$, either all correct measurers agree on the same probabilistic maximum inter-region latency $\tdecide$, or at least one faulty node or faulty link will be detected; and (2) after the fault is detected, a new latency can be decided within bounded time, all correct measurers and log keepers will decide on the same latency, and the decided value is from a valid proposal.}

\begin{proof}
    For (1): Consider a region $R_i$. All correct measurers send their accept messages at $t_{n}^{\mathsf{acc}}$, which should arrive at every node in $R_i$ by $t_{n}^{\mathsf{dec}}$ (see Fig.~\ref{fig:timeline}). Since the value accepted by each measurer is the minimum of all proposed values plus $\Dinter$, if all measurers behave correctly, the accepted messages will have matching latency value and hence all nodes will agree/decide on the same value. 
    If a properly signed accept message does not arrive at a node $N_y$ by $t_{n}^{\mathsf{dec}}$, $N_y$ will declare an intra-region omission fault against the sender; and if an accept message contains a different value, $N_y$ will also detect the discrepancy. 

   For (2):  We have shown in Appendix~\ref{subsec:app-detection-recovery-meas} that when a fault is detected at $t_{\mathsf{dclr}}$, a new latency can be decided by $t_{\mathsf{dclr}} + 5\tintra +  E_{\mathsf{dclr\_v}} + E_{\mathsf{log\_ex}} + E_{\mathsf{log\_v}} + E_{\mathsf{decide}}$.   During stage 4 in the recovery process, all new accept messages are forwarded to all nodes, and thus any equivocation will be detected. Therefore, all nodes have the same view of the new accept messages. Following a deterministic process, they will all decide on the same value.
    For a latency to be decided by a correct node, the proposal of that latency must exist in the log of $f_i+1$ nodes. A correct node will not record an invalid proposal in its log. Therefore, the proposal must be valid and thus the latency is from a valid proposal. 
\end{proof}

\section{Proof of detection and recovery properties}
\subsection{Handling commission faults: Consensus on message correctness}\label{sec:appendix-consensus-poc}
\noindent{\bf Theorem~\ref{thm:consensus-poc} (Consensus on correctness).}
{
\em      All correct nodes in a downstream region $R_j$ agree on 1) whether an input message $m$ from an upstream region $R_i$ is correct, and 2) whether safe mode should be triggered. 
}
\begin{proof}
    If a correct replica in $R_j$ deems $m$ correct, it must have received $m$ and a valid PoC for $m$, and must have forwarded both to all peer replicas. Since the intra-region network in $R_j$ is reliable, all correct replicas will receive the valid PoC and deem $m$ correct as well.

    Suppose a correct node determines that the safe mode should be triggered. Then, either a correct input message $m$ for one of its tasks, or a heartbeat for the current round must have failed to arrive in time. In the former case, no correct replica could have received a valid PoC for $m$, as a correct replica always forwards a valid PoC for each correct input message it receives. Therefore, they will all decide to trigger the safe mode. In the latter case, we have shown in Theorem~\ref{thm:agreement} that all correct measurers will agree on the same $\taccept$, which is broadcast in the region, and a timeout will be triggered only if $\taccept$ is the indicator for timeout. As correct nodes have the same $\taccept$, they will  agree on triggering the safe mode.
\end{proof}

\subsection{Handling omission faults: TGS properties}\label{sec:appendix-tgs-properties}
\noindent{\bf Theorem~\ref{thm:TGS}-\ref{item:good-score}. }
    {\em  Let $E(N)$ be node $N$'s expected score change in each
        round. For all pairs of correct nodes $(N_s, N_r)$, where $N_s$ is the sender and $N_r$ is the receiver of inter-region messages, $E(N_s) > 0$ and $E(N_r) > 0$.}

\begin{proof}
    Let $(N_s, N_r)$ be a pair of correct nodes. Then, the probability of them having a suspicious behavior is smaller than $1 - \pnorm$. Recall that a node's score is decreased (penalized) by $\spenaty$ whenever it exhibits a suspicious behavior, and increased (awarded) by  $\saward$ whenever it exhibits a normal behavior (i.e., sends or receives an inter-region message in time). Therefore, the score change expectation for $N_s$ and $N_r$ is   
    $E(N_s) = E(N_r) > \saward\cdot \pnorm - \spenaty\cdot (1-\pnorm) = \saward\cdot \pnorm \cdot (1-\alpha) > 0$.
\end{proof}

\vspace{0.5ex}
\noindent {\bf Theorem~\ref{thm:TGS}-\ref{item:long-term-bound}} 
    {\em To keep a positive score, a node
        must behave normally for no less than $p'= \frac{\alpha \cdot
        \pnorm}{1+(\alpha-1)\pnorm}$ of the inter-region messages in the long term.}

\begin{proof}
    Suppose a node $N$ drops or delays $1-p'$ of the inter-region messages and delivers the rest normally. Then, the expectation of score change for $N$ is
    $E(N) = \saward\cdot p' - \spenaty \cdot (1-p') $. 
    Since the initial score for $N$ is positive ($s_{\mathsf{init}} = 1$), the node needs $E(N) \geq 0$ to maintain a positive score in the long term. In other words, 
    \begin{align*}
        E(N) \geq 0 &\Leftrightarrow p' \geq \frac{\spenaty}{\spenaty + \saward} \\
    &\Leftrightarrow p'\geq \frac{\alpha \cdot \pnorm}{1+(\alpha-1)\pnorm} 
    \end{align*}
Hence, the theorem.
\end{proof}

\vspace{0.5ex}
\noindent {\bf Theorem~\ref{thm:TGS}-\ref{item:short-term-bound}.}
{\em  To keep a positive score,
        within any time window of size $w = \beta + k + k\cdot \alpha \cdot \frac{\pnorm}{1-\pnorm}$,
        a node must behave normally for more than $w-(\beta + k)$ messages ($\forall k \in \mathbb{N}^*$)
}
\begin{proof}
    Consider any window of size $w = \beta + k + k\cdot \alpha \cdot \frac{\pnorm}{1-\pnorm}$. Suppose a node $N$ delivers no more than $k\cdot \alpha \cdot\frac{\pnorm}{1-\pnorm}$ of the inter-region messages in time while having a positive score. Then, at least $\beta+k$ messages must have been delayed or dropped. 
    By definition, $\spenaty / \saward = \alpha \cdot \pnorm / (1-\pnorm)$, and thus 
    $$k\cdot \alpha\cdot\frac{\pnorm}{1-\pnorm} =\frac{k\cdot\spenaty}{\saward}. $$
    As a result, the total score awarded during this window will be at most 
    $\frac{k\cdot\spenaty}{\saward} \cdot \saward = k\cdot \spenaty$. Similarly, the total score penalized will be at least $ (\beta+k)  \cdot \spenaty = s_{\max}+k\cdot\spenaty$ (since $\spenaty= s_{\max} / \beta$). Hence, the score change of the node $N$ during the window $w$ will be no more than $k\cdot s_{pen} - (s_{\max}+k\cdot\spenaty) = -s_{\max}$, which implies $N$'s score is negative. Thus, the theorem holds by contradiction.
\end{proof}

\subsection{Recovery propagation: BTR guarantee}\label{sec:appendix-btr-property}
\noindent{\bf Theorem~\ref{thm:rt-prop} (BTR Guarantee).}
{
\em       If a fault is detected at $\tdetect$, then all nodes in the system that need to perform recovery will start the recovery or switch to a safe mode by
     $\tdetect+D_{\mathsf{RP}}$ where 
     $D_{\mathsf{RP}} = 
     2 \cdot (\Delta_{\mathsf{det}}+\tinterval+ 2\tintra + E_{\mathsf{hb}} + \thbtimeout)
     $
     if no further fault occurs in $[\tdetect, \tdetect+D_{\mathsf{RP}}]$.   
}

\begin{proof}
    We first show that all nodes involved can start recovery after at most two hops of RP. If the faulty node $N_y$ in $R_j$ is detected by a node in the same region, nodes in $R_j$ will perform local recovery and propagate it to the region(s) that $N_y$ communicates with. This involves one hop of RP. If the detector is in a different region $R_i$ instead, then nodes in $R_i$ should propagate the recovery to $R_j$, and nodes in $R_j$ further propagate it to the region(s) that communicates with any task in $R_j$ that is assigned a new replica during $R_j$'s recovery. This makes a two-hop RP.

    For each hop, consider an RP from $R_i$ to $R_j$. A fault detected at $t_{\mathsf{det}}$ in $R_i$ takes at most $\tintra$ to reach a measurer $N_x\in R_i$. 
    After receiving the fault declaration, $N_x$ waits at most $\Delta_{\mathsf{det}}+\tinterval$ until the signature exchange phase of round $n$ starts ($t_n^s$). Here, the first quantify is to ensure all measurers are aware of the fault (since $\Delta_{\mathsf{det}}$ is the maximum difference in the time to detect the fault at different nodes), and the second quantity is the maximum waiting time until the next heartbeat round (since $\tinterval$ is the heartbeat period).
    By the time $t_{n}^s + \tintra + E_{\mathsf{hb}} + \thbtimeout = t_{n}+\thbtimeout$ (time for measurers in $R_j$ to accept the latency), the measurers in $R_j$ will either have received the heartbeats with verifiable $m_{\mathsf{rp}}$, or trigger a safe mode. This one-hop RP takes no more than $(\tintra + \Delta_{\mathsf{det}}+ \tinterval) + (\tintra + E_{\mathsf{hb}} + \thbtimeout)$. Thus, a two-hop RP takes at most $2\cdot (\Delta_{\mathsf{det}}+ \tinterval + 2\tintra + E_{\mathsf{hb}} + \thbtimeout)$.
\end{proof}

\end{document}